\documentclass[aip,prl,reprint]{revtex4-2} 
\usepackage[T1]{fontenc}
\usepackage{times}
\usepackage{graphicx}

\usepackage{amsfonts, amsmath,amssymb, bm, graphicx}
\usepackage{siunitx}

\usepackage[english]{babel}
\usepackage{lipsum}

\def\vec#1{{\bm{#1}}}
\def\mat#1{{\hat{\vec{#1}}}}

\newcommand{\figref}[2]{\hyperref[#1]{\ref{#1}(#2)}}
\usepackage{physics}
\usepackage{lipsum}
\usepackage{changes}

\usepackage[colorlinks=true,allcolors=blue]{hyperref}
\usepackage{footmisc}

\usepackage{upgreek}

\usepackage{soul}

\begin{document}

{
\makeatletter
\def\frontmatter@thefootnote{%
 \altaffilletter@sw{\@fnsymbol}{\@fnsymbol}{\csname c@\@mpfn\endcsname}%
}%

\makeatother
\title{Curvilinear spin-wave dynamics beyond the thin-shell approximation:\\ Magnetic nanotubes as a case study}

\author{L. K\"orber}\email{l.koerber@hzdr.de}
\affiliation{Helmholtz-Zentrum Dresden - Rossendorf, Institut f\"ur Ionenstrahlphysik und Materialforschung, D-01328 Dresden, Germany}
\affiliation{Fakultät Physik, Technische Universit\"at Dresden, D-01062 Dresden, Germany}

\author{R. Verba}
\affiliation{Institute of Magnetism, Kyiv 03142, Ukraine}

\author{Jorge A. Ot\'alora}
\affiliation{Departamento de Física, Universidad Católica del Norte, Avenida Angamos 0610, Casilla 1280, Antofagasta, Chile}

\author{V. Kravchuk}
\affiliation{Institut für Theoretische Festk\"orperphysik
Karlsruher Institut für Technologie, 76131 Karlsruhe, Germany}
\affiliation{Bogolyubov Institute for Theoretical Physics of the National Academy of Sciences of Ukraine, 03143 Kyiv, Ukraine}
%\author{G. Quasebarth}
%\affiliation{Helmholtz-Zentrum Dresden - Rossendorf, Institut f\"ur Ionenstrahlphysik und Materialforschung, D-01328 Dresden, Germany}
%\affiliation{Fakultät Physik, Technische Universit\"at Dresden, D-01062 Dresden, Germany}

%\author{A. Otto}
%\affiliation{Fakultät Physik, Technische Universit\"at Dresden, D-01062 Dresden, Germany}

%\affiliation{Helmholtz-Zentrum Dresden - Rossendorf, Institut f\"ur Ionenstrahlphysik und Materialforschung, D-01328 Dresden, Germany}
%\affiliation{Fakultät Physik, Technische Universit\"at Dresden, D-01062 Dresden, Germany}

\author{J. Lindner}
\affiliation{Helmholtz-Zentrum Dresden - Rossendorf, Institut f\"ur Ionenstrahlphysik und Materialforschung, D-01328 Dresden, Germany}

\author{J. Fassbender}
\affiliation{Helmholtz-Zentrum Dresden - Rossendorf, Institut f\"ur Ionenstrahlphysik und Materialforschung, D-01328 Dresden, Germany}
\affiliation{Fakultät Physik, Technische Universit\"at Dresden, D-01062 Dresden, Germany}

\author{A. Kákay}%\email{a.kakay@hzdr.de}
\affiliation{Helmholtz-Zentrum Dresden - Rossendorf, Institut f\"ur Ionenstrahlphysik und Materialforschung, D-01328 Dresden, Germany}

\date{\today}

\begin{abstract}
Surface curvature of magnetic systems can lead to many static and dynamic effects which are not present in flat systems of the same material. These emergent magnetochiral effects can lead to frequency nonreciprocity of spin waves, which has been shown to be a bulk effect of dipolar origin and is related to a curvature-induced symmetry breaking in the magnetic volume charges. So far, such effects have been investigated theoretically mostly for thin shells, where the spatial profiles of the spin waves can be assumed to be homogeneous along the thickness. Here, using a finite-element dynamic-matrix approach, we investigate the transition of the spin-wave spectrum from thin to thick curvilinear shells, at the example of magnetic nanotubes in the vortex state. With increasing thickness, we observe the appearance of higher-order radial modes which are strongly hybridized and resemble the perpendicular-standing-waves (PSSWs) in flat films. Along with an increasing dispersion asymmetry, we uncover the curvature-induced non-reciprocity of the mode profiles. This is explained in a very simple picture general for thick curvilinear shells, considering the inhomogeneity of the emergent geometric volume charges along the thickness of the shell. Such curvature-induced mode-profile asymmetry also leads to non-reciprocal hybridization which can facilitate unidirectional spin-wave propagation. With that, we also show how curvature allows for nonlinear three-wave splitting of a higher-order radial mode into secondary modes which can also propagate unidirectionally. We believe that our study provides a significant contribution to the understanding of the spin-wave dynamics in curvilinear magnetic systems, but also advertises these for novel magnonic applications.
\end{abstract}

\maketitle

\section{Introduction}\label{sec:intro}

During the last years, three-dimensional architectures became the focus of interest in several research areas such as the ferromagnets and superconductors. One of the most promising explored effects are related to the bending of the samples into curved shells, with the bending radius being comparable with the characteristic length scales of the studied system, dependent on the underlying order parameter and interactions. As shown in various theoretical papers, in ferromagnetic systems surface curvature of the samples can lead to emergent anisotropies~\cite{gaidideiCurvatureEffectsThin2014a,shekaCurvatureEffectsStatics2015,shekaTorsioninducedEffectsMagnetic2015a,korniienkoCurvatureInducedMagnonic2019} as well as emergent magnetochiral interactions.\cite{bordacsChiralityMatterShows2012a,hertelCurvatureInducedMagnetochirality2013, kezsmarkiOnewayTransparencyFourcoloured2014a,gaidideiCurvatureEffectsThin2014a,shekaCurvatureEffectsStatics2015,shekaNonlocalChiralSymmetry2020} This, for example, can result in curvature-induced stabilization of Skyrmions and Merons on Gaussian and paraboloid bumps\cite{kravchukMultipletSkyrmionStates2018,eliasWindingNumberSelection2019}, pinning of domain walls~\cite{lewisMagneticDomainWall2009,yershovCurvatureInducedDomain2015,volkovExperimentalObservationExchangeDriven2019}, localization of magnon modes in curved magnetic nanowires~\cite{gaidideiLocalizationMagnonModes2018}, a magnon band structure for a nanowires with periodically deformed shape\cite{korniienkoCurvatureInducedMagnonic2019} or an asymmetric spin-wave dispersion in magnetic nanotubes\cite{otaloraCurvatureInducedAsymmetricSpinWave2016,otaloraAsymmetricSpinwaveDispersion2017}. 
Many effects have been reported, that can lead to non-reciprocal spin-wave propagation in ferromagnetic samples. These are all related to some kind of a symmetry breaking, as for example in the case of the Dzyaloshinskii-Moriya interaction (DMI)~\cite{cortes-ortunoInfluenceDzyaloshinskiiMoriya2013,moonSpinwavePropagationPresence2013} the symmetry is inheritable broken by the asymmetric nature of the interaction itself. In the case of the dipolar interaction, the non-reciprocal spin-wave propagation is induced by the symmetry breaking in the magnetic volume pseudo-charges, as shown in many exciting works.~\cite{mikaDipolarSpinwaveModes1985,grunbergWaysModifySpin1985,grunbergLayeredMagneticStructures1986,gladiiFrequencyNonreciprocitySurface2016,otaloraCurvatureInducedAsymmetricSpinWave2016,slukaEmissionPropagation1D2019a,gallardoReconfigurableSpinWaveNonreciprocity2019a,ishibashiSwitchableGiantNonreciprocal2020,albisettiOpticallyInspiredNanomagnonics2020,grassiSlowWaveBasedNanomagnonicDiode2020,gallardoSpinwaveFocusingInduced2021} In order for this magneto-dipolar symmetry breaking to happen (in media with homogeneous material parameters) the spin-wave propagation direction and the magnetic equilibrium magnetization have to satisfy a certain geometry. Loosely speaking, the system formed by the wave vector of the spin wave $\bm{k}$ and the magnetic equilibrium state $\bm{m}_0$ within the cross section perpendicular to the propagation direction has to exhibit some sense of chirality. This is the case, for example, for the magnetostatic surface spin waves in antiferromagnetically coupled bilayers,\cite{slukaEmissionPropagation1D2019a,gallardoReconfigurableSpinWaveNonreciprocity2019a,ishibashiSwitchableGiantNonreciprocal2020,albisettiOpticallyInspiredNanomagnonics2020,gallardoSpinwaveFocusingInduced2021} when the propagation direction is perpendicular to the magnetization in the layers [see Fig.~\figref{fig:FIG1}{a}], or the spin waves propagating along Bloch walls in thin magnetic media with perpendicular-to-plane magnetic anisotropy [see Fig.~\figref{fig:FIG1}{b}].\cite{zhangFerromagneticDomainWall2018,henryUnidirectionalSpinwaveChanneling2019} In both cases the system of $\bm{m}_0$ and $\bm{k}$ exhibit a handedness and the spin-wave propagation is nonreciprocal due to dynamic dipolar fields. In this simple picture, it should be clear that, for the spin waves along N\'eel walls in thin easy-plane ferromagnets, such kind of nonreciprocity does not take place.\cite{korberSpinWaveReciprocityPresence2017}

\begin{figure}[h!]
    \centering
    \includegraphics{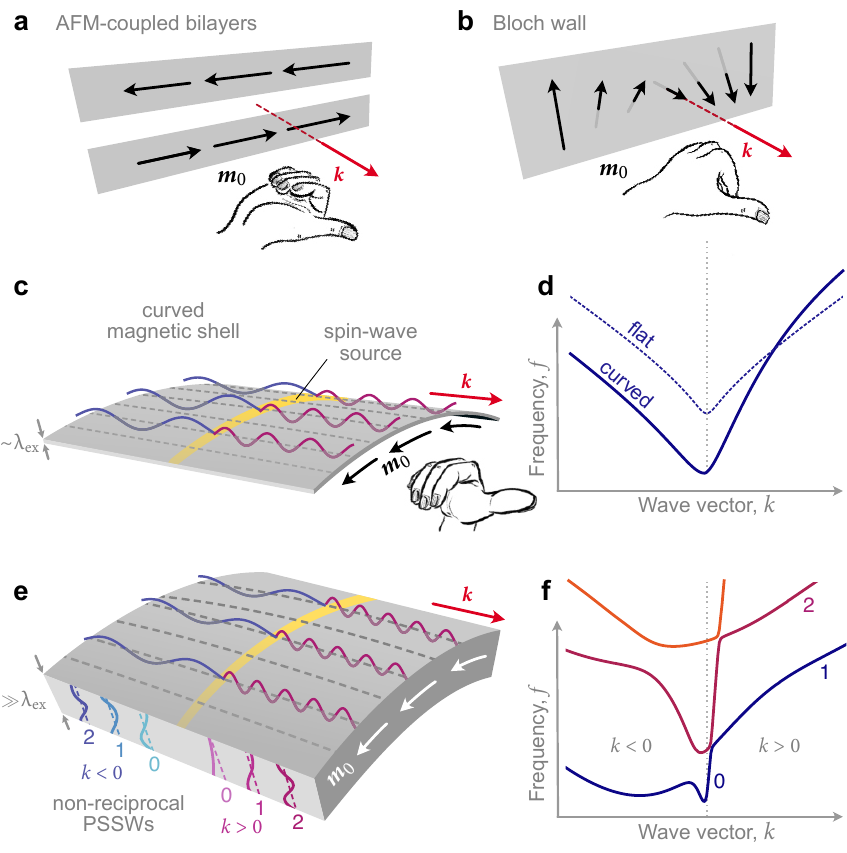}
    \caption{In panels (a) and (b) the sketch of an antiferromagnetically coupled bilayers and that of a Bloch wall, together with the spin wave propagation direction is shown. (c) Sketch of a curved magnetic shell magnetized along one of the principal curvature axis and of the spin waves excited with a spin-wave source, propagating perpendicular to the ground magnetic state ($\bm{m}_0$). The counter propagating waves have different wavelengths. The $\lambda_{ex}$ is the characteristic magnetic length. (d) Comparison of the dispersion relation for spin waves in flat and curved magnetic thin films. The curvature induced magnetochiral effects result in an asymmetric dispersion. (e) Sketch of a curved magnetic film with thickness larger than $\lambda_{ex}$ and the same excitation geometry as in (c). Similar to flat thick films, the modes will be confined along the thickness, thus forming PSSW modes. Due to the curvature, the counter propagating modes will have non-reciprocal mode profiles. This will result in a complex spin-wave dispersion, with strong asymmetry and avoided level crossings due to mode hybridization. An exemplary dispersion for thick magnetic tubes is shown in (f).}
    \label{fig:FIG1}
\end{figure}

We note that the above-mentioned symmetry breaking can be formulated in a more exact and general manner using the toroidal moment $\bm{\tau}=\sum_i \bm{r}_i\times\bm{s}_i$, which is the sum of the individual cross products of the magnetic spins. Namely, {if the wave vector has a non-zero projection on the toroidal moment, $\bm{\tau}\cdot\bm{k}\neq 0$, non-reciprocal spin-wave propagation is allowed by symmetry considerations, since the symmetry operations will transform both quantities the same way.\cite{spaldinToroidalMomentCondensedmatter2008a,kezsmarkiOnewayTransparencyFourcoloured2014d,kezsmarkiEnhancedDirectionalDichroism2011b,szallerSymmetryConditionsNonreciprocal2013b,okamuraMicrowaveMagnetoelectricEffect2013,kocsisIdentificationAntiferromagneticDomains2018} Of course, the underlaying interactions will then play an important role for the actual appearance of the non-reciprocal spin-wave propagation.

In curvilinear magnetism, magnetochiral symmetry breaking of dipolar origin can be introduced by the surface curvature of a magnetic shell, as shown in Fig.~\figref{fig:FIG1}{d}. As has been reported before for thin magnetic nanotubes\cite{otaloraCurvatureInducedAsymmetricSpinWave2016,otaloraAsymmetricSpinwaveDispersion2017}, bending a shell along the magnetization direction $\bm{m}_0$ will lead to dispersion asymmetry for the spin waves with a wave vector component perpendicular to $\bm{m}_0$, as shown in Fig.~\figref{fig:FIG1}{d}. This concept of curvature-induced magneto-dipolar symmetry breaking has been generalized theoretically by Sheka \textit{et al.},\cite{shekaNonlocalChiralSymmetry2020} where the authors discuss the important role of the emergent geometric contribution to the magnetic volume charges, highlighting that this is clearly a bulk effect. However, up to now, curvilinear magnetism has been described theoretically only for thin shells within the framework of continuum theory where the magnetization (statics and dynamics) are assumed to be homogeneous along the thickness of the curved shells. This approximation is valid as long as the thickness does not exceed the order of magnitude of the exchange length $\lambda_\mathrm{ex}$ of the material. 

In this paper we show that the spin-wave spectrum changes quite drastically when transitioning to thick shells [Fig.~\figref{fig:FIG1}{e}], for which the magnetic oscillations can be inhomogeneous along the thickness. Our calculations are carried out for the example of vortex-state nanotubes of increasing thickness, for which we calculate the spin-wave dispersion and mode profiles using our recently developed finite-element dynamic-matrix approach for propagating spin waves.\cite{korberFiniteelementDynamicmatrixApproach2021} Many of the observed phemona and the related discussion, however, also hold for general thick magnetic shells. Note, that we discuss shells which are extrusions along the normal direction of a curved surface, such that the curvature along the thickness of the shell is well defined and identical with the local curvature of each extruded surface. We show, that naturally, with increasing thickness, the spin-wave spectrum becomes much denser and we observe the appearance of strongly hybridized higher-order radial modes which resemble the perpendicular standing spin-waves (PSSWs) in flat films or general curved shells [Fig.~\figref{fig:FIG1}{e}]. However, in addition to an increase in the curvature-induced dispersion asymmetry (sketched in Fig.~\figref{fig:FIG1}{e,f}), we also find a curvature-induced non-reciprocity of the mode profiles along the thickness. In a simple picture, we explain this by the inhomogeneity of the emergent geometric volume charges along the thickness of the shell. We discuss that this curvature-induced mode-profile asymmetry naturally leads to non-reciprocal dipole-dipole hybridization, which, in return, can be exploited to facilitate unidirectional spin-wave propagation. For spin waves under resonant (linear) excitation we show how this can be used to construct a magnonic diode using a single magnetic nanotube. Finally, we also discuss the influence of shell curvature on nonlinear spin-wave interaction in thick shells, where, in particular, we show
that magneto-dipolar symmetry breaking in nanotubes allows
for three-wave splitting of higher-order radial modes into azimuthal modes which can also propagate unidirectionally.

We believe that this work provides a considerable contribution in the fundamental understanding of magnetization dynamics in curvilinear shells, advancing theoretical considerations and understanding beyond the thin-shell approximation. Moreover, the emerging dynamic effects could be of interest for modern magnonic applications, such as spin-wave diodes or nonlinear magnonic circuits.

\section{Methodology}

In this section we will introduce the studied system, that is a magnetic nanotube with a fixed average radius and varying shell thickness and introduce how we numerically calculate the spin-wave normal modes.

\subsection{Studied magnetic system}

For our study of curvilinear spin-wave dynamics in thick magnetic shells, we select the case of magnetic nanotubes in the flux-closure/vortex state, $\bm{m}_0 = \bm{e}_\varphi$ (see Fig.~\figref{fig:FIG2}{a}), for which a curvature-induced spin-wave dispersion asymmetry has been predicted by Otálora \textit{et al.} in 2016 and which (in the case of thin shells) have been studied extensively in the literature.\cite{otaloraCurvatureInducedAsymmetricSpinWave2016, otaloraAsymmetricSpinwaveDispersion2017,otaloraFrequencyLinewidthDecay2018,salazar-cardonaNonreciprocitySpinWaves2021} For the case of thin-shell tubes, where the thickness of the nanotube mantel is in the order of the magnetic exchange length $\lambda_\mathrm{ex}$, the spatial profiles of the spin waves propagating along such tubes are proportional to $\exp[i(kz+m\varphi)]$. The modes are characterized by their wave vector $k$ along the axis of the tube (here, the $z$ axis) and an integer index $m$ which can take positive and negative values and counts the number of oscillation periods in the azimuthal ($\varphi$) direction. Again, due to the strong influence of the exchange interaction, the mode profiles can be taken as homogeneous along the radial direction. Naturally, when increasing the nanotube thickness, we will need a further mode index to characterize the radial mode profiles.

As examples, in Fig.~\figref{fig:FIG2}{b}, we show the lateral mode profiles of the spin-wave modes in a nanotube with \SI{65}{\nano\meter} average radius and \SI{10}{\nano\meter} mantel thickness at $k=0$ up to azimuthal index $m=\pm 4$. It is important to note that, in the vortex state, the modes with opposite azimuthal index $\pm m$ form degenerate doublets, $\omega_m(k) = \omega_{-m}(k)$,\cite{otaloraCurvatureInducedAsymmetricSpinWave2016, otaloraAsymmetricSpinwaveDispersion2017,otaloraFrequencyLinewidthDecay2018,salazar-cardonaNonreciprocitySpinWaves2021} while curvature-induced dispersion asymmetry appears in the $z$ direction, \textit{i.e.} $\omega_m(k) \neq \omega_m(-k)$. This is, indeed, in line with the simple chirality consideration made in Sec.~\ref{sec:intro}. 

\begin{figure}[h!]
    \centering
    \includegraphics{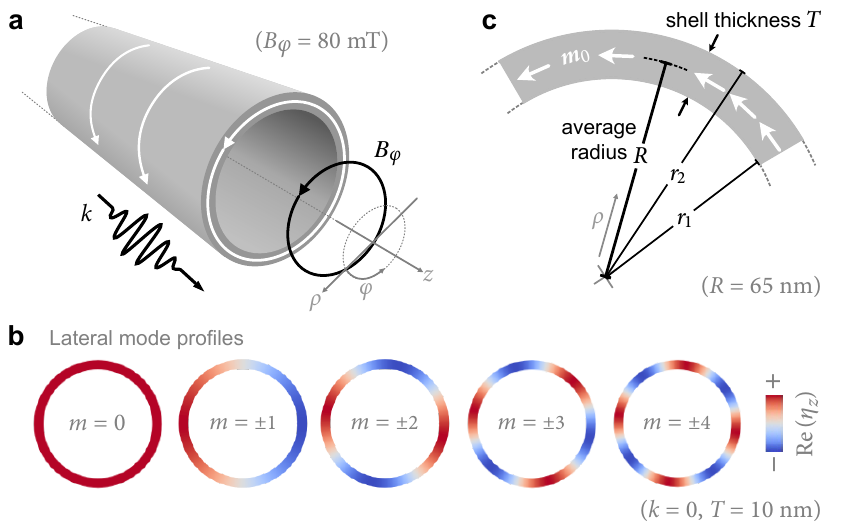}
    \caption{(a) Sketch of a thin-shell magnetic nanotube with $R=\SI{65}{\nano\meter}$ radius and $T=\SI{10}{\nano\meter}$ mantel thickness in a vortex magnetic state, stabilized by an azimuthal field. The propagartion direction of spin waves is along the length of the nanotube, namely the z direction. To study the effect of the shell thickness on the spin-wave dispersion, the average radius ($R=\SI{65}{\nano\meter}$) of the tube is kept fixed and the inner and outer radii are varied, as shown in panel (b). For thin shells the mode profiles are translation invariant along the thickness of the tube. In panel (c) exemplary mode profiles are shown, starting with the homogeneous mode ($m=0$) up to the $m=\pm4$, with $m$ being the azimuthal mode index.}
    \label{fig:FIG2}
\end{figure}

We consider typical material parameters of the common soft magnetic alloy Ni$_{80}$Fe$_{20}$ (permalloy). In particular, we assert a saturation magnetization of $\mu_0 M_\mathrm{S}=\SI{1}{\tesla}$, an exchange-stiffness constant of $A_\mathrm{ex}=\SI{13}{\pico\joule/\meter}$ and a reduced gyromagnetic ratio of $\gamma/2\pi = \SI{28}{\giga\hertz/\tesla}$. For nanotubes with a finite length, the vortex state (which is necessary for dipole-induced dispersion asymmetry) naturally appears for tubes with large enough average radius $R$ and aspect ratio $R/L$, with $L$ being the length of the tube.\cite{landerosEquilibriumStatesVortex2009} For quasi-infinite tubes, however, the vortex state is unstable due to the cost in exchange energy provided by the curling of the magnetization along the azimuthal direction. The exchange field arising from this curling can be compensated by a magneto-crystalline easy-plane anisotropy or by the application of an external magnetic field.\cite{otaloraOerstedFieldAssisted2015,salazar-cardonaNonreciprocitySpinWaves2021} For simplicity, we investigate the case of zero magneto-crystalline anisotropy, $K=0$, and stabilize the vortex state using an external field in $\varphi$ direction of $B_\varphi=\SI{80}{\milli\tesla}$. 
In order to disentangle thickness effects from spurious changes in curvature when changing the dimensions of the nanotube, in this study, we keep the average radius $R=(r_1+r_2)/2$ of the tube (and therefore, the average curvature radius) constant at $R=\SI{65}{\nano\meter}$ and vary only the thickness $T$ of the shell (see Fig.~\figref{fig:FIG2}{c}) in a range between 10 and \SI{90}{\nano\meter}. In terms of the average curvature radius $R$ and the shell thickness $T$, the critical external field to stabilize a vortex state is given by
\begin{equation}\label{eq:critical-field}
    B_{\varphi,\mathrm{crit}} =  \frac{\mu_0 M_\mathrm{s}\lambda_\mathrm{ex}^2}{TR}\ln\qty(\frac{2R+T}{2R-T})
\end{equation}
which, at constant average radius, increases monotonically with the shell thickness.\footnote{The critical field in Eq.~\ref{eq:critical-field} can be obtained from Ref.~\citenum{salazar-cardonaNonreciprocitySpinWaves2021} by replacing inner and outer radii $r_{1,2}$ with thickness and average radius as $r_{1,2}=R\mp T/2$.} However, for the given thickness range and material parameters, this critical field remains below \SI{10}{\milli\tesla}.

\subsection{Calculation of spin-wave normal modes}

\begin{figure*}[t!]
    \centering
    \includegraphics{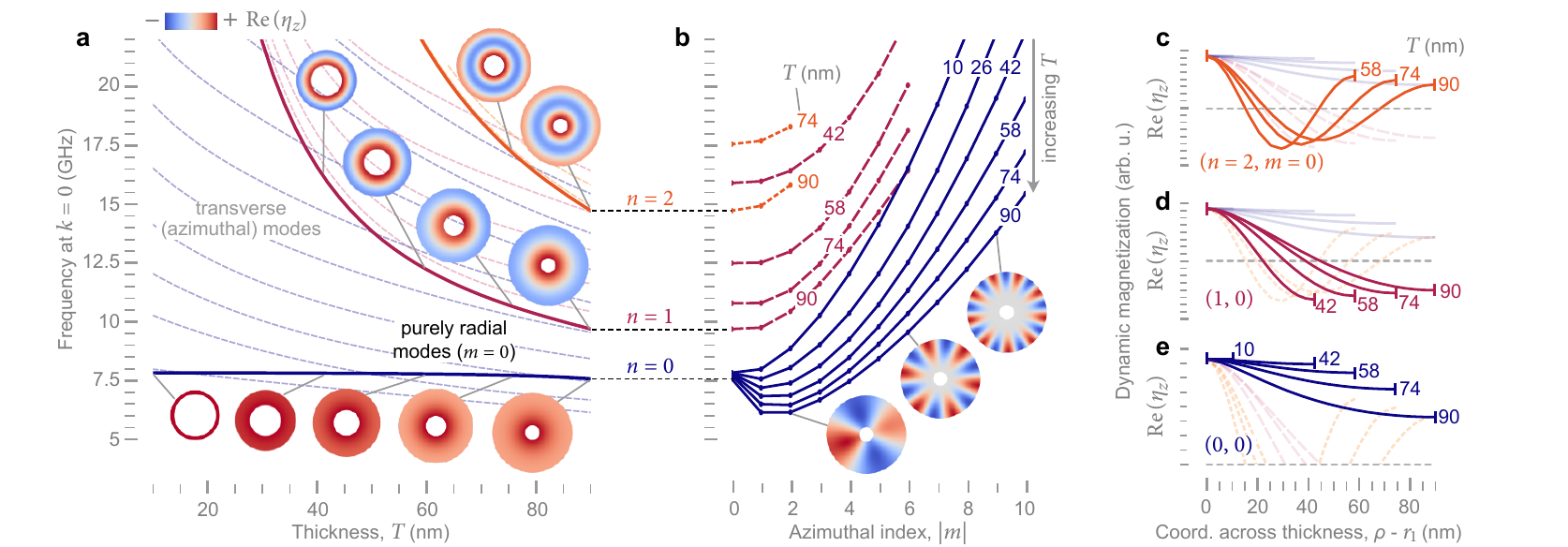}
    \caption{The frequency evolution of the modes at $k=0$ with an emphasis on the first three purely radial modes ($m=0$) in function of the tube thickness is summarized in (a). As expected, with increasing thickness the frequency of the higher order radial modes decreases. (b) The frequency evolution of the modes for different tube thicknesses versus the azimuthal mode index. In this case the propagation is parallel to $\bm{m}_0$, namely in Backward-volume-magnetostatic-wave (BVMSW) geometry. With increasing thickness the BVMSW-like character (negative group velocity) of the dispersion increases. Modes with a higher azimuthal index are localized closer to the outer mantel. This is attributed to the exchange interaction. Panels (c-e) show the line-cut of the mode profiles along the nanotube thickness for several thicknesses and three different radial indices, $n=0$, 1 and 2 and $m=0$ azimuthal index. Note, the modes are still unpinned, therefore for thick thickness range the exchange boundary conditions are still dominant.}
    \label{fig:FIG3}
\end{figure*}
Even though, for thin vortex-state tubes and for thick nanowires in the axial state (magnetized along the $z$ direction), the spin-wave spectrum has already been described theoretically, an analytical consideration of the spin-wave spectrum in thick vortex-state nanotubes is extremely challenging. Although, the vortex ground state exhibits a cylindrical symmetry, neither the exchange-only approximation nor the dipolar-only approximation would reduce the linearized Landau-Lifshitz equation of motion of the magnetization to a Bessel form. 

Here, instead, we calculate the spin-wave normal modes in thick vortex-state nanotubes numerically using our recently developed finite-element dynamic-matrix approach for propagating spin waves in waveguides with arbitrarily-shaped cross section\cite{korberFiniteelementDynamicmatrixApproach2021} which is implemented in the open-source micromagnetic-modeling package \textsc{TetraX}.\cite{korberTetraXFiniteElementMicromagneticModeling2022} In this method, the linearized equation of motion of the (unitless) dynamic magnetization $\delta \bm{m}(\bm{r},t)$ is converted to an eigenvalue problem which is solved for general propagating spin waves of the form
\begin{equation}
    \delta \bm{m}(\bm{r},t) \propto \bm{\eta}_{\nu,k}(x,y)e^{ikz}e^{-i\omega_\nu(k)t}
\end{equation}
with $\nu$ being some mode index, $\bm{\eta}_{\nu,k}$ being the (complex-valued) lateral mode profiles, $k$ being the wave vector along the $z$ direction and $\omega_\nu(k)$  being the corresponding dispersion. The problem is simplified by projecting the equation of motion into a single cross section of the tube and calculating only for the lateral mode profiles $\bm{\eta}_{\nu,k}$ and corresponding frequencies $\omega_\nu(k)$ for each wave vector $k$.

%profiles of the form $\bm{m}_\nu = \bm{\eta}_{\nu,k}\exp(ikz)$ by projecting the eigenvalue problem Eq.~\eqref{eq:ev-problem} into a single cross section of the waveguide and solving for the lateral mode profiles $\bm{\eta}_{\nu,k} \equiv \bm{\eta}_{\nu,k}(x,y)$. In particular, the magnetic tensors are transformed according to 
%
%\begin{equation}
%    \vu{N}_k = e^{-ikz}\vu{N}e^{ikz}
%\end{equation}
%
%which allows to write Eq.~\eqref{eq:ev-problem} as a wave-vector dependent eigenvalue problem for the lateral mode profile only,
%
%\begin{equation}
%    \frac{\omega_\nu(k)}{\omega_M} \bm{\eta}_{\nu,k} = \vu{D}_k \bm{\eta}_{\nu,k} \quad \text{with}\quad \bm{\eta}_{\nu,k} \perp \bm{m}_0.
%\end{equation}
%
%with $\vu{D}_k$ being a wave-vector-dependent dynamic matrix. 

For our case we only consider exchange, dipolar and Zeemann interaction. We employ standard exchange boundary conditions $\bm{n}\cdot\bm{m}=0$ (with $\bm{n}$ being the normal direction) at the boundaries of the cross section.
The lateral dipolar potential of the spin waves is calculated using the hybrid finite-element/boundary-element method by Fredkin and Koehler\cite{fredkinHybridMethodComputing1990} which has recently been extended to solve the screened Poisson equation of propagating waves.\cite{korberFiniteelementDynamicmatrixApproach2021} 
%To implement the constraint $\bm{\eta}_{\nu,k} \perp \bm{m}_0$, the eigenvalue problem is projected into the subspace locally orthogonal to the equilibrium magnetization $\bm{m}_0$.\cite{korberFiniteelementDynamicmatrixApproach2021a}
All involved operators and vector fields are discretized using the finite-element method (FEM) while subdividing the cross sections of the different nanotubes into triangles. The characteristic discretization length of the meshes is continuously varied between \SI{3}{\nano\meter} (for the thinnest tube, $T=\SI{10}{\nano\meter}$) and \SI{6}{\nano\meter} (for the thickest tube, $T=\SI{90}{\nano\meter}$) in order to optimize computational time. The resulting discretized sparse linear system is numerically diagonalized using an iterative Arnoldi-L\'{a}nczos method.\cite{lanczosIterationMethodSolution1950,arnoldi1951principle} A major benefit of the propagating-wave dynamic-matrix is that it directly yields the spin-wave dispersion $\omega_\nu(k)$ and the corresponding lateral mode profiles $\bm{\eta}_{\nu,k}$ without any additional post processing necessary as is the case for micromagnetic simulations, which rely on a time integration of the nonlinear Landau-Lifshitz-Gilbert equation of motion. Moreover, by modeling the different nanotubes only within a single cross section, the computational cost is drastically reduced, leading to a computational time for the largest shell thickness in the order of one hour, which would take several days using standard time-domain micromagnetic simulations.\cite{korberFiniteelementDynamicmatrixApproach2021} This allows us to study the smooth transition of the spin-wave dispersion in the given thickness range in steps of \SI{2}{\nano\meter}. For each thickness $T$, the lowest 100 modes/branches were calculated for a range of 201 wave vectors between $k_\mathrm{min}=\SI{-40}{\radian/\micro\meter}$ and $k_\mathrm{max}=\SI{+40}{\radian/\micro\meter}$.

\section{Results and Discussion}

\subsection{Thickness dependence of normal-mode frequencies and profiles}

In this section, we investigate the change in the overall spatial mode profiles and frequencies when increasing the shell thickness, starting from the well-known thin-shell case. 

\subsubsection{Non-propagating modes at $k=0$: Appearance of higher-order radial modes}\label{sec:radial-modesk0}

Before presenting the thickness-dependence of the full spin-wave dispersion, as a first step, we restrict ourselves to the modes at $k=0$, which exhibit a homogeneous profile along the nanotube axis and could be measured using standard ferromagnetic-resonance (FMR) experiments. In Fig.~\figref{fig:FIG3}{a}, we show the smooth transition of the oscillations frequencies $f=\omega/2\pi$ at $k=0$ as the thickness of the nanotube shell is increased from \SI{10}{\nano\meter} to \SI{90}{\nano\meter}. To ease the visualization, for now, we focus only on the purely radial modes ($m=0$) while the azimuthal modes ($m\neq 0$) are only shown in the background. We see that, with increasing shell thickness, the overall frequencies decrease and additional modes descent into the lower-frequency range. The insets in Fig.~\figref{fig:FIG3}{a}, which show the corresponding $z$ components of the dynamical magnetization (the real part of the $z$ component of the complex lateral mode profiles $\bm{\eta}$), reveal that these additional modes correspond to higher-order radial modes that exhibit a standing-wave character along the radial ($\rho$) direction. The modes for $k=0$ are similar to those known from vortex disks, with well-defined radial and azimuthal mode indices.\cite{ivanovMagnonModesThin2002,buessFourierTransformImaging2004a,zaspelExcitationsVortexstatePermalloy2005} In a general curvilinear shell, these modes correspond to the perpendicular-standing spin-wave (PSSW) modes of flat magnetic films. For very thin shells, however, these modes are at very high frequencies due to the cost in exchange energy related to necessarily high wave vectors in radial/out-of-plane direction. As for the PSSWs in thin films or the radial modes in magnetic disks, we denote these modes by an additional mode index $n = 0, 1,2,...$ and so forth. At $k=0$, this index is identical with the number of nodal lines along the radial/out-of-plane direction. However, we will soon see that this association of the radial index with the number of nodal lines is, in general, not valid for the whole wave vector space, because modes with different number of nodal lines across the radius will be shown to be strongly hybridized. Moreover, we will also see that the curvature of the magnetic system can induce a change in the number of nodal lines for certain wave vectors $k$. These effects, however, are not present at $k=0$, thus we can safely associate $n$ with the number of radial nodal lines.

Naturally, with increasing shell thickness, the influence of the dipolar interaction on the spin-wave dynamics, in particular, on the mode frequencies, becomes more important. To highlight this, in Fig.~\figref{fig:FIG3}{b},  we change visualization and show the same mode frequencies at $k=0$ for selected thicknesses of $T=10$, 26, 42, 58, 74 and \SI{90}{\nano\meter} as a function of the (magnitude of the) azimuthal index $\abs{m}$. Recall, that the tubes are magnetized in the vortex state $\bm{m}_0=\bm{e}_\varphi$. As a result, with increasing $\abs{m}$ the angular component of the total spin-wave wave vector, $k_\varphi = m/R$, parallel to the equilibrium magnetization $\bm{m}_0$, increases. For tubes with a thin shell ($T=\SI{10}{\nano\meter}$), the dependence of the frequency with respect to the angular wave vector is still quadratic, $f(k=0) \propto k_\varphi^2$, and, therefore, dominated by the exchange interaction.\cite{otaloraCurvatureInducedAsymmetricSpinWave2016} However, for thicker tubes ($T\gtrsim\SI{40}{\nano\meter}$), the dispersion in $\varphi$ direction deviates from this behavior and develops a minimum at small azimuthal wave vectors. In flat films, this is known as backward-volume-wave (BVMSW) behavior, where spin waves propagating parallel to the equilibrium magnetization exhibit negative group velocities at small wave vectors. This dipolar effect increases with film thickness and originates from a decrease of the total magnetic energy (and therefore, spin-wave frequency) by bringing magnetic surface charges closer to each other (increasing $k_\varphi$), and therefore, reducing stray fields. To no surprise, the same frequency-azimuthal-index relation is also observed in flat vortex-state disks with large enough radius.\cite{gaidideiMultipleVortexantivortexPair2010,schultheissExcitationWhisperingGallery,korberNonlocalStimulationThreemagnon2020,verbaTheoryThreemagnonInteraction2021}

In parallel, with increasing $\abs{m}$, we observe a localization of the spin waves to the outer mantel of the tube (see the insets in Fig.~\figref{fig:FIG3}{b}), which can be attributed to savings in exchange energy and has also been observed before in micrometer-sized disks for very large azimuthal indices.\cite{schultheissExcitationWhisperingGallery}

In Figs.~\figref{fig:FIG3}{c-d}, we show the $z$ components of the radial mode profiles only for the different radial modes $n=0,1,2$ and for different thicknesses $T$ across the thickness of the shell. It is possible to see that, even for $T=\SI{90}{\nano\meter}$, the modes remain mostly unpinned at the boundaries of the cross section.

\begin{figure*}[t!]
    \centering
    \includegraphics{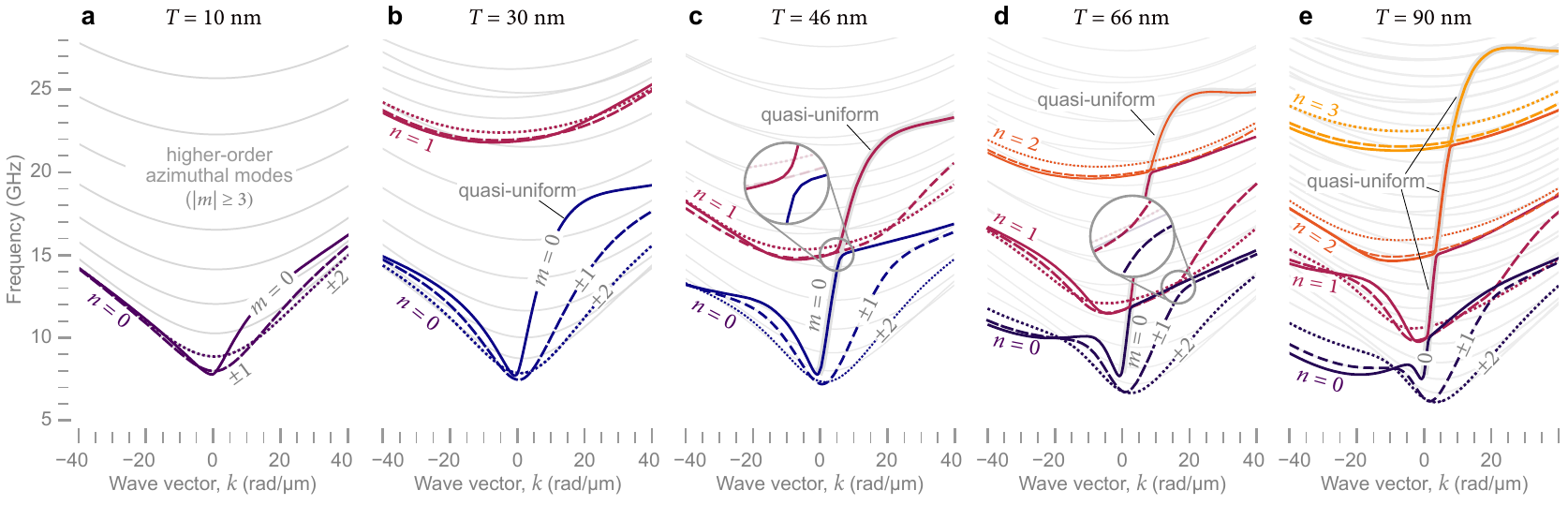}
    \caption{In (a-e) we show the transition of the dispersion as a function of the tube thickness for five exemplary thicknesses in the vortex state for modes propagating along the nanotube axis. The evolution of three modes $m=0, \pm 1, \pm 2$ for each presented thickness is highlighted. With increasing thickness the overall asymmetry of the dispersion increases. Already above $T=\SI{30}{nm}$ the higher order radial modes appear in the displayed frequency range. When branches of modes with the same $m$ but different $n$ are approaching, due to the dipole-dipole interaction hybridization will take place. This will show up in the dispersion in the form of avoided level crossings. Such avoided level crossing are magnified in panel (c) for $m=0$ modes and in panel (d) for $m=\pm 1$ modes.}
    \label{fig:FIG4}
\end{figure*}
\subsubsection{Dispersion of propagating modes}

After having discussed the thickness evolution of the modes at $k=0$ and having observed the emergence of higher-order radial modes, we would now like to investigate how the full dispersion $\omega_{nm}(k)$ of the different modes evolves when increasing the shell thickness $T$. For this, in Fig.~\ref{fig:FIG4}, we show the dispersion of all modes, in a given frequency range, for five exemplary  thicknesses of $T=10$, 30, 46, 66 and \SI{90}{\nano\meter}. For visual clarity, again, we highlight only the lowest-azimuthal modes $m=0,\pm 1,\pm 2$, while all higher-order azimuthal modes $\abs{m}\geq 3$ are drawn with light gray lines. Recall, that modes with the same wave vector $k$ and radial index $n$ but opposite azimuthal index $\pm m$ are degenerate. As we do not change the symmetry of the system when increasing the shell thickness $T$, this degeneracy remains also for thick tubes. However, with increasing thickness, as expected, the overall mode frequencies decrease and the spectrum becomes much denser. As the first-order radial modes ($n=1$) approach the zeroth-order modes ($n=0$), we observe that different radial branches exhibit avoided level crossings with each other, indicating the fact that they are hybridized. This is nicely seen for example in Fig.~\figref{fig:FIG4}{c} for a thickness of $T=\SI{46}{\nano\meter}$ where the mode ($n=0, m=0$) shares an avoided level crossing with the mode ($n=1, m=0$). As we increase the thickness further, we see that the quasi-uniform mode (zero nodal lines across the radius) \textit{pinches} through all the other radial branches, leading to an enormous increase in dispersion asymmetry. It can also be seen, that for $k<0$, level repulsion seems to be much larger than for $k>0$. To investigate this hybridization more closely, in the next section, we will directly analyze the lateral mode profiles. Moreover, we refer to the supplementary material, which shows the transition of the $m=0$ modes, and how they hybridize, for even more thicknesses. We have also attached a movie showing the continuous transition of all modes with increasing thickness in steps of \SI{2}{\nano\meter}. 

At this point it becomes already clear that a categorization of the radial modes by their number of nodal lines across the radial direction is not suitable, as the number of nodal lines of a mode can change when varying the wave vector $k$. In other words: Modes with a well-defined number of radial nodal lines are not actual normal modes. Instead, the true normal modes are mixed by dynamic dipolar fields, an effect being commonly referred to as dipole-dipole hybridization. This hybridization is quite ubiquitous in spin-wave dynamics and also appears, for example, for the PSSWs in flat films,\cite{qinExchangetorqueinducedExcitationPerpendicular2018,tacchiStronglyHybridizedDipoleexchange2019} between the standing-wave modes across the width of transversally magnetized rectangular waveguides\cite{korberNumericalReverseEngineering2021} or between the radial modes in axially magnetized cylindrical nanowires,\cite{rychlySpinWaveModes2019} just to name a few. We will see soon, however, that in contrast to the previous examples, the dipole-dipole hybridization in thick vortex-state nanotubes is strongly perturbed by the curvature-induced magneto-chirality of the system. 

For completeness, it is important to note that only modes with the same azimuthal index $m$ are hybridized. In Fig.~\figref{fig:FIG4}{d}, we can see another example, where only the different radial modes with $m=1$ or $m=-1$ are hybridized. Loosely speaking, whether two \textit{bare} spin-wave modes can hybridize via dipolar interaction depends crucially on their spatial overlap and, thus, on the symmetry of their mode profiles. In cylindrical systems, azimuthal modes with different indices $m\neq m^\prime$ belong to different irreducible representations of the symmetry point group defined by the ground state, and share zero spatial overlap. Such a hybridization could, however, be introduced by lowering the continuous rotational symmetry.\cite{korberModeSplittingSpin2022}

As mentioned  before, the asymmetry of the dispersion, \textit{i.e.}, the frequency nonreciprocity of the spin-waves in vortex state nanotubes strongly increases with increasing shell thickness. We note that this increase in asymmetry $\Delta f = f(k) - f(-k)$ is already predicted by the analytic thin-shell theory for nanotubes by Otálora \textit{et al.},\cite{otaloraCurvatureInducedAsymmetricSpinWave2016, otaloraAsymmetricSpinwaveDispersion2017,otaloraFrequencyLinewidthDecay2018,salazar-cardonaNonreciprocitySpinWaves2021} (see supplementary material) due to an increased symmetry breaking in the radially-averaged magnetic volume charges. We will come back to the role of these volume charges later. \textit{Nota bene}, the thin-shell theory does assume homogeneous mode profiles along the thickness and, of course, cannot take any inhomogeneity of the mode profiles along the thickness into account. Following, we will see that this inhomogeneity of the mode profiles along the thickness direction will contribute to the frequency non-reciprocity. In fact, we will see that the mode profiles will become non-reciprocal themselves.

\subsection{Curvature-induced nonreciprocity of mode profiles}

\begin{figure*}[t!]
    \centering
    \includegraphics{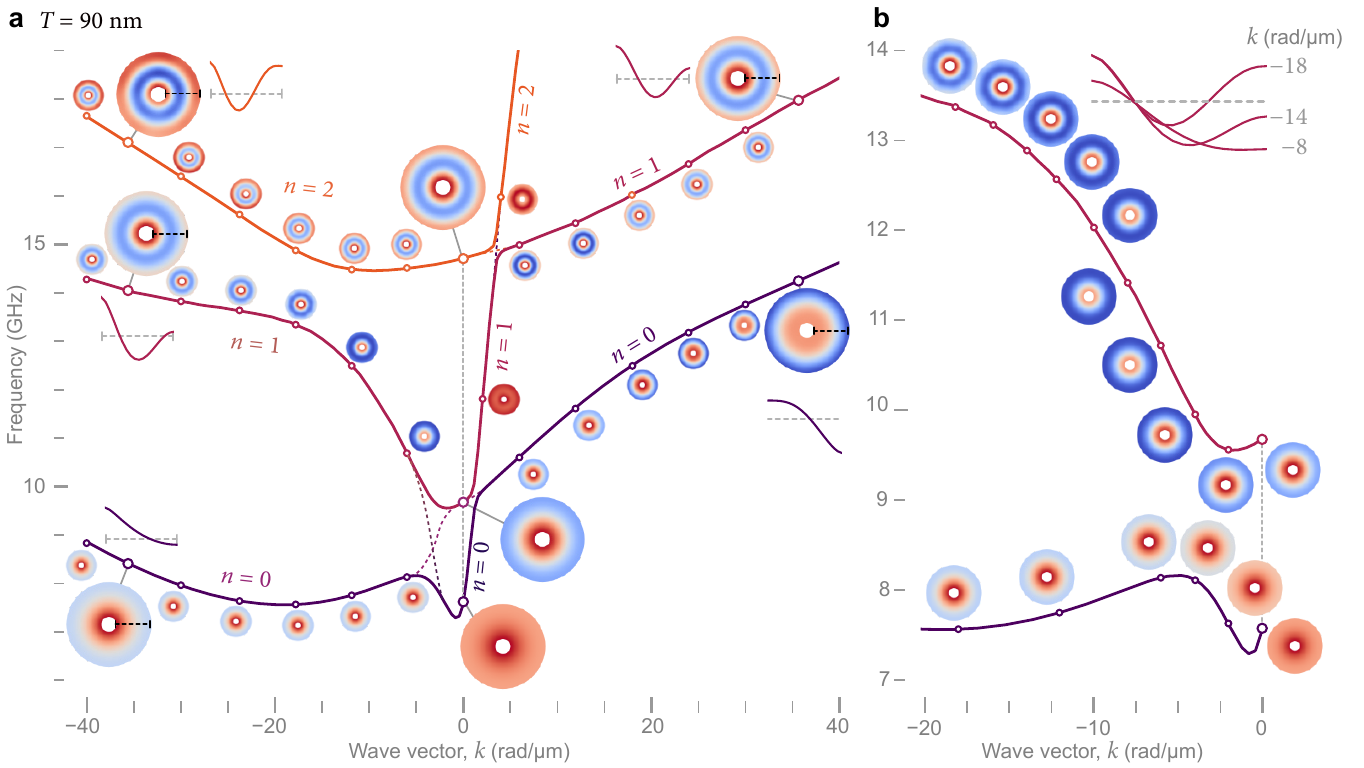}
    \caption{Dispersion of the first three radial modes  with $m=0$ is shown in (a) together with the radial modes profiles for selected wave vectors. For five particular modes at $k=\pm \SI{36}{\radian/\micro\meter}$ the radial mode profile along a cut-line is also attached. For positive $k$ values the evolution of the mode profiles after avoided level crossings is more or less as expected. However, for negative $k$ values the quasi-uniform mode disappears and large avoided level crossings are forming between the branches, indicating a strong interaction between them. In panel (b) we highlight the mode profile evolution of the first two radial branches. As seen on the lower branch, after the avoided level crossing the uniform-like mode transforms into a mode with a nodal line along its radial direction. On the higher branch the mode with a single nodal line will however transform into a mode with two nodal lines for larger negative $k$ values. This process is also indicated by the inset showing line profiles for three different $k$ values.}
    \label{fig:FIG5}
\end{figure*}

In this section we discuss the radial mode profiles, their change due to the hybridization and the observed nonreciprocity for different spin-wave branches. We will show, that the mode-profile nonreciprocity originates from the dynamic dipolar fields created by the magnetic charges, namely the appearance of the geometric charge due to the non-zero mean curvature of the nanotube. Since the geometrical charge is present for any general curved geometry with non-zero mean curvature, its effect, namely the mode-profile nonreciprocity is not only present for tubular geometries but for curved samples in general.

\subsubsection{Nonreciprocity of radial profiles in tubes}

As we have seen in the previous section, increasing the tube thickness leads to an enhanced asymmetry of the overall dispersion, compared to the thin shell tube, and avoided level crossings appear for multiple branches. These level crossings between different radial modes of the same azimuthal index $m$ appear to be qualitatively very different for the opposite propagation directions ($\pm k$). Again, as the degree of dipole-dipole hybridization between different modes depends on their spatial overlap, this is a strong hint that the mode profiles themselves are nonreciprocal. To shed light on this, in Fig.~\ref{fig:FIG5}, the first three radial modes with $m=0$ azimuthal mode index for the thickest considered nanotube ($T=\SI{90}{\nano\meter}$) are shown together with the lateral mode profiles at selected $k$ values. Moreover, at $k=\pm\SI{36}{\radian/\micro\meter}$, we show line cuts of the lateral profiles along the radius (the radial profiles) as insets. To guide the eye of the reader, in Fig.~\figref{fig:FIG5}{a}, we have added dashed lines in the avoided level crossings.   
Starting from the modes at $k=0$, which we have discussed in Sec.~\ref{sec:radial-modesk0}, let us first analyze the spin waves propagating with positive wave vectors $k>0$. Note, that the role of positive/negative wave vectors is reversed when changing the vortex circularity from clockwise to counter-clockwise or vice versa. The lowest branch closed to $k=0$ is the quasi-uniform mode which already at small wave vectors hybridizes with the first radial mode. According to the expectations, after the avoided level crossing the mode profile of the lowest branch is changed, resembling the mode profile of the second branch. Thus the quasi-uniform mode will be higher in frequency after the avoided level crossing. The same scenario is repeated when the quasi-uniform mode hybridizes with the mode of the third branch with two radial nodal lines. As a result, the quasi-uniform mode will have a steep increase in frequency. Both frequency gaps at the avoided level crossings are small, suggesting a weak hybridization of the modes between the branches. Overall, it is possible to see that, up to weak hybridization and other slight modifications, the structure of the radial mode profiles is preserved. Thus, for $k>0$, and far from any avoided level crossings, it is still reasonable to talk about the radial modes in terms of their number of radial nodal lines.

This situation changes quite drastically for the opposite propagation direction $k<0$. We see that, at small negative wave vectors, the radial profile of the zeroth-order mode $n=0$ (quasi uniform) transforms into the one of the first-order mode $n=1$ (one nodal line). At the same time, however, the mode $n=1$ does \textit{not} transform into a quasi-uniform mode. Instead, it acquires first one, and then two nodal lines for even more negative $k$ values and far from this avoided level crossing. In fact, at $k=-\SI{36}{\radian/\micro\meter}$, we have one mode with one nodal line and two modes with two nodal lines. We show this apparent \textit{disappearance} of the uniform precession for negative $k$ even more closely in Fig.~\figref{fig:FIG5}{b}. Of course, the situation here is not quite trivial, since we observe two concurrent effects. On one hand, we see that there is an intrinsic non-reciprocity in the radial mode profiles which leads to the disappearance of the quasi-uniform mode. On the other side, this perturbation of the mode profiles allows for extremely strong coupling, and therefore, leads to very large avoided level crossings between the modes. It is clear, that the non-reciprocity of the mode profiles has to be a dipolar effect. In thin vortex-state tubes, exchange interaction is not able to induce any asymmetry\cite{otaloraCurvatureInducedAsymmetricSpinWave2016} and, qualitatively, this interaction does not change when increasing the tube thickness. It is also clear, that for $k<0$, the usual categorization of radial modes by their number of nodal lines completely loses its meaning. 

We note that the strong influence of the dipolar interaction on the mode profiles makes the standard approach to calculate the \textit{unhybridized} dipole-exchange spectrum in thin films unusable. In flat geometries, analytic expressions for the dipole-exchange spectrum can usually be obtained by first obtaining the exchange mode profiles, projecting the linearized equation of motion in these profiles and then setting all inter-mode-coupling terms nil.\cite{kalinikosTheoryDipoleexchangeSpin1986,gladiiFrequencyNonreciprocitySurface2016,tacchiStronglyHybridizedDipoleexchange2019,grassiSlowWaveBasedNanomagnonicDiode2020} Using numerically obtained exchange profiles, in the supplementary material, we show that this perturbation theory fails for thick nanotubes. Furthermore, assuming unperturbed (exchange) profiles, which are completely reciprocal, the dispersion asymmetry is much weaker.

\subsubsection{Origin of mode-profile nonreciprocity in general thick curved shells}

The origin of the nonreciprocity of the radial modes in vortex-state nanotubes can be understood in a very simple picture which is general for the Damon-Eshbach waves (with $\bm{k}\perp\bm{m}_0$) in magnetic shells with finite mean curvature and with their equilibrium magnetization aligned along a principle curvature direction. The crucial point here is the interaction between the magnetic surface charges $\sigma = \bm{n}\cdot\bm{m}$ (with the outer shell normal $\bm{n}$) and the volume charges $\varrho = -\nabla\bm{m}$ which are the sources of the dynamic dipolar fields realizing a curvature-induced chiral symmetry breaking.
Within the curvilinear frame of reference attached to the geometry of a curved shell, the volume charges $\varrho$ are not only given by the spatial derivatives of the magnetization but also by a term which is proportional to the mean curvature of the magnetic shell. For a (generalized) cylindrical surface of arbitrary shape one has 
\begin{equation}\label{eq:volume-charges}
    \varrho(\bm{r}) = -\nabla\bm{m}(\bm{r})= \varrho_\mathrm{int} \underbrace{- \mathcal{H}(\bm{r})m_n(\bm{r})}_{\varrho_\mathrm{g}}.
\end{equation}
Here, $m_n$ is the normal component of the magnetization and $\mathcal{H}(\bm{r}) = \mathcal{H}_0/(1+\zeta \mathcal{H}_0)$ is the local mean curvature of the shell, $\mathcal{H}_0$ is the mean curvature of the central surface and $\zeta$ is the coordinate along the normal direction (for the central surface $\zeta=0$).\footnote{Note, that the mean curvature $ \mathcal{H}(\bm{r})$ is also defined along the thickness because of that, as we discuss here only shells which consist of extrusions of curved surfaces along their normal direction. Thus the local curvature inside the volume of the shell is given by the surface curvature of the respective extruded surface.} The first term in Eq.~\eqref{eq:volume-charges} is denoted here as the \textit{intrinsic} charge $\varrho_\mathrm{int}$ which arises due to the spatial variation of the magnetization,
\begin{equation}
    \varrho_\mathrm{int}= -\Bigg(\frac{1}{1+\zeta\mathcal{H}_0} \partial_1 m^1 + \partial_2 m^2 + \partial_\zeta m_n\Bigg).
\end{equation}
Here, the indices 1 and 2 correspond to coordinates (magnetization components) along the cylinder directrix and generatrix, respectively.
The second term in Eq.~\eqref{eq:volume-charges} is the \textit{geometric} (or extrinsic) charge $\varrho_\mathrm{g}$. The concept of this geometric charge was first introduced by Sheka \textit{et al.} in Ref.~\citenum{shekaNonlocalChiralSymmetry2020} where the authors discuss the curvature-induced chiral symmetry breaking in thin shells. In their work, the authors refer to the first term in Eq.~\eqref{eq:volume-charges} only as the \textit{tangential} charges, as the magnetization is taken to be homogeneous along the thickness and the intrinsic charges are only given by the covariant derivatives along the surface of the plane. However, as we show here, this concept can also be used to explain the curvature effects in thick shells as well. Of course, in the case of thick shells, one has to take into account also the spatial derivatives of the magnetization along the thickness. As a result, here we use the term \textit{intrinsic} charge.

\begin{figure}[h!]
    \centering
    \includegraphics{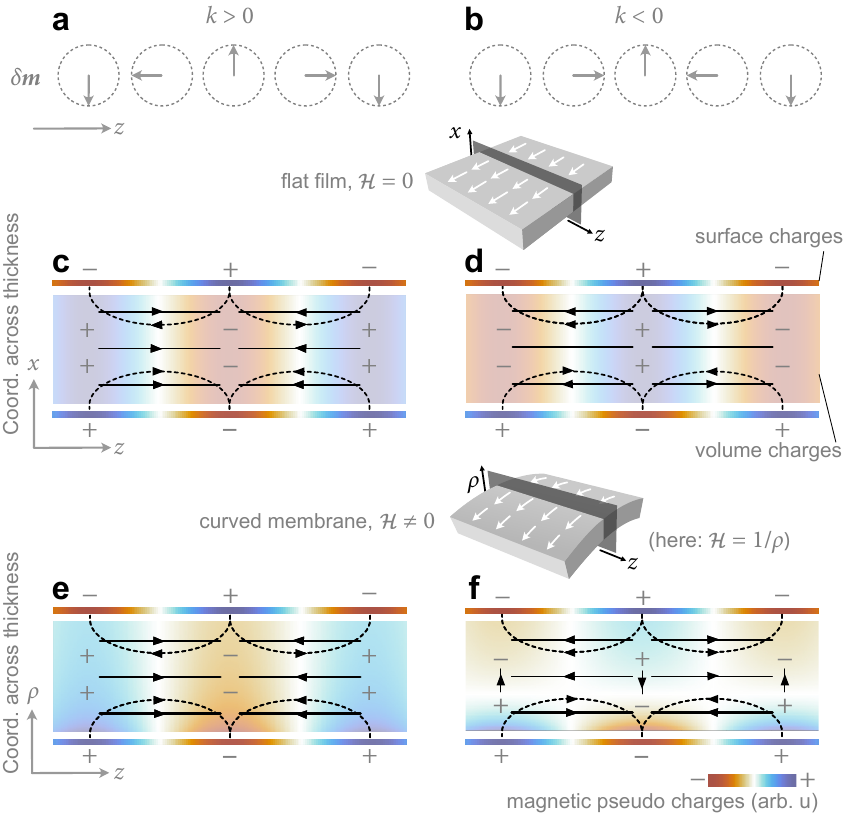}
    \caption{Sketch of spin waves propagating in positive as well as to negative $k$ direction are shown in panel(a) and (b), respectively. The arrows denote the dynamic component only along the propagation direction. The volume and surface charge distribution for a flat thick film, as a result of the spatial dependence of the magnetization dynamics together with the resulting demagnetizing dynamic fields are depicted in (c) and (d) according to the propagation directions in (a) and (b). The sign change in $k$ simply results in a sign change of the magnetic charges. The magnetic charge distribution for thick curved shells together with the dynamic dipolar fields is shown in panels (e) and (d). For $k>0$, the geometric charge $g = -\mathcal{H}m_\rho$ (with the mean curvature $\mathcal{H}=1/\rho$) contributes to the regular volume charge (intrinsic charge) such, that the resulting dipolar field is in-line to that for flat films. By changing the propagation direction, the sum of the intrinsic and geometrical charges will produce a nodal line,  resulting in mode profile asymmetry and disappearance of the uniform mode.}
    \label{fig:FIG6}
\end{figure}

For comparison, let us quickly review the situation in flat films for which it is already well known that the dynamic dipolar fields can lead to a localization of the spin waves propagating with $\bm{k}\perp\bm{m}_0$ to either surface of the film, depending on the propagation direction. In particular, we consider a \textit{trial} wave with a homogeneous profile along the thickness, propagating in positive $z$ direction ($k>0$) in a film magnetized along the positive $y$ direction ($\bm{m}_0 = \bm{e}_y$). In Fig.~\figref{fig:FIG6}{a}, we show the corresponding dynamic magnetization $\delta \bm{m}=[\cos(kz), 0, \sin(kz)]$ at a fixed point in time along the propagation direction. In Fig.~\figref{fig:FIG6}{c}, we schematically show the corresponding surface- and volume charges (heat maps), and the field lines of the resulting dipolar fields generated by such a wave. For flat systems, obviously $\mathcal{H}=0$, and the intrinsic charges are synonymous with the volume charges, $\varrho = -k\cos(kz)$. The surface charges are simply given by $\sigma = \pm\cos(kz)$, depending on the surface. As can be seen in Fig.~\figref{fig:FIG6}{c}, at one surface, the dipolar fields generated by the volume and by the surface charges compensate each other, whereas at the other surface they aggregate. As a consequence, a mode with homogeneous profile along the thickness does not represent an allowed normal mode of the magnetic system. In reality, to compensate this field asymmetry and to minimize internal dipolar fields, the true mode will be localized to one surface of the film and exponentially decay into the volume. As the propagation direction is reversed [Figs.~\figref{fig:FIG6}{b,d}], only the volume charges change in sign and a localization to the opposite surface is favored. This localization to opposite surfaces depending on the propagation direction can already be exploited to introduce a frequency nonreciprocity for spin waves in completely flat systems using different surface anisotropies on either surface.\cite{gladiiFrequencyNonreciprocitySurface2016} However, in homogeneous materials, a reversal of the propagation direction will only lead to a mirroring of the mode profile about the film plane. 

In curved samples, however, this mirror symmetry is broken by the geometrical charges $\varrho_\mathrm{g}$. Considering our case of a thick tube in the vortex state ($\bm{m}_0 = \bm{e}_\varphi$), or even only the case of a transversally magnetized tube segment, the mean curvature is given by $\mathcal{H}=1/\rho$. As a result, the volume charges of a mode which is homogeneous along the thickness are given by
\begin{equation}
    \varrho = \underbrace{-k\cos(kz)}_{\varrho_\mathrm{int}}  \underbrace{- \frac{\cos(kz)}{\rho}}_{\varrho_\mathrm{g}}.
\end{equation}
Here the components of $\delta\vec{m}=[\cos(kz), 0, \sin(kz)]$ aer defined in the curvilinear basis. Clearly, even for a homogeneous mode profile along the thickness, the curvature of a magnetic shell already introduces an inhomogeneity of the volume charges along the thickness of the shell which, unlike the intrinsic charge $ -k\cos(kz)$, is independent of the propagation direction. In Fig.~\figref{fig:FIG6}{e}, we see that, for $k>0$, this leads to an increase of the volume charges at the inner surface of the shell. However, the overall situation is the same as in a flat film, only with slight modifications. This is in agreement with the previous observations we made for the nanotube with $T=\SI{90}{\nano\meter}$ in Fig.~\figref{fig:FIG5}{a}, where the modes for $k>0$, in general, preserved the same structure as for $k=0$. For the opposite propagation direction, $k<0$, the situation is drastically different, as the geometric and intrinsic charges can now compensate each other. In the range $-1/r_1 < k < -1/r_2$ (with $r_1$ and $r_2$, again, being inner and outer curvature radius of the shell), this compensation can even lead to a nodal line of the volume charges at a certain position across the shell thickness, as depicted in Fig.~\figref{fig:FIG6}{f}. This, in return, can not only lead to the situation that the dipolar fields from surface and volume charges aggregate everywhere, but it can totally change the landscape of the dynamic dipolar fields generated by the mode. As a result, the mode profiles for $k<0$ are heavily perturbed by the inhomogeneous compensation of intrinsic and geometrical charges. This can even lead in a thick nanotube, as seen in the previous section, to the transformation of the zeroth-order mode (or quasi-uniform mode) into a mode with nodal lines. It is clear that this curvature-induced dipolar non-reciprocity of the mode profiles is a bulk effect like the non-reciprocity in magnetically inhomogeneous media\cite{gallardoSpinwaveNonreciprocityMagnetizationgraded2019, grassiSlowWaveBasedNanomagnonicDiode2020}, but unlike the mode profile asymmetry in the previously mentioned systems with asymmetric surface anisotropy\cite{gladiiFrequencyNonreciprocitySurface2016} or systems with interfacial DMI,\cite{kostylevInterfaceBoundaryConditions2014} which are surface effects.

We note, that the emergence of a geometrical charge in curvilinear shells is also the origin of the frequency non-reciprocity itself, as already reported for thin-shell nanotubes previously.\cite{otaloraCurvatureInducedAsymmetricSpinWave2016} However, in the case of thin shells, an inhomogeneity (and possible non-reciprocity) of the mode profiles along the thickness is, of course, suppressed by the influence of the exchange interaction.

\section{Consequences for linear and nonlinear spin-wave dynamics}

Finally, we wish to examine some of the consequences of the curvature-induced symmetry breaking for the linear and nonlinear spin-wave dynamics in thick magnetic shells, demonstrating on thick magnetic nanotubes. As an example for the linear spin-wave dynamics, we show how the curvature-induced mode-profile asymmetry and its consequence, the non-reciprocal dipole-dipole hybridization, can be used to achieve unidirectional spin-wave propagation and therefore, construct a magnonic diode. As an outlook for curvature effects on nonlinear magnetization dynamics, finally, we investigate the effect of the curvature on the three-magnon splitting of higher-order radial modes in magnetic nanotubes.

\subsection{Unidirectional propagation and diode behavior}

\begin{figure}[h!]
    \centering
    \includegraphics{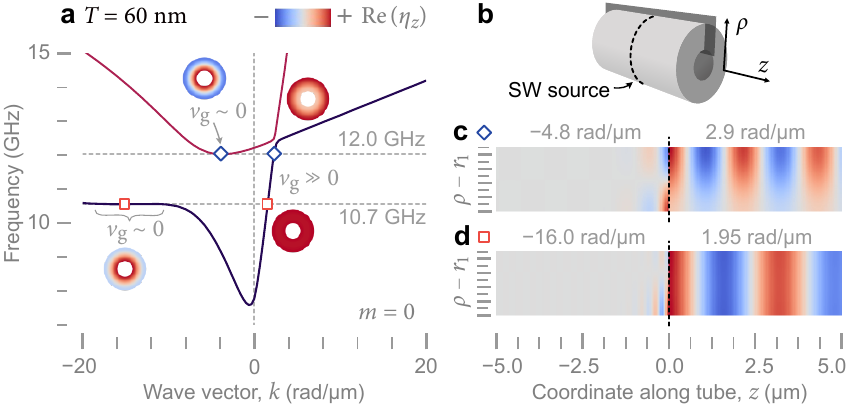}
    \caption{The dispersion of two hybridized branches for the \SI{60}{nm} thick nanotube is shown in (a). For particular frequency values, namely \SI{10.7}{GHz} and \SI{12}{GHz}, marked with grey dashed lines, the spin-wave transport is unidirectional, as further indicated with the close to zero group velocity of the modes for negative and large group velocities for positive wave vectors. The mode profiles at these frequencies are included as little insets, coloured by the z component of the dynamic magnetization. (b) Sketch of a nanotube cross section with a spin-wave source in the middle. The modes excited with this source at \SI{10.7}{GHz} and \SI{12}{GHz} are shown in (c) and (d) using profiles $\bm{\eta}\exp(ikz-z/L)$ with the attenuation length $L=v_g/\Gamma$. Note, the propagation is decaying fast for negative $k$ values and the profiles are nonreciprocal along the radius}
    \label{fig:FIG7}
\end{figure}

Within the previous sections, we have seen that the curvature-induced non-reciprocity in the radial mode profiles naturally leads to a non-reciprocal dipole-dipole hybridization of different radial modes (with the same azimuthal index $m$). In particular, modes with $k<0$ are hybridized much stronger than for $k>0$ which leads to a significantly larger gap at the avoided level crossings. Inspired by a recent work of Grassi \textit{et al.},\cite{grassiSlowWaveBasedNanomagnonicDiode2020} who proposed and implemented the same idea using a magnetic bilayer system, this hybridization asymmetry can be used to construct a slow-wave-based magnonic diode, in which spin waves propagate only in one direction. In our case, the diode behavior is achieved simply by tuning the thickness of the nanotube. In Fig.~\figref{fig:FIG7}{a}, we see that, for example, for a thickness of $T=\SI{60}{\nano\meter}$ the curvature-induced mode-profile asymmetry produces a large avoided level crossing for $k<0$ between two pure radial modes ($m=0$), which, in return, results in a plateau of the dispersion, where the frequency remains nearly constant around \SI{10.7}{\giga\hertz} over a wide wave-vector range. In this range, the group velocity of the spin waves, $v_\mathrm{g}=\partial\omega/\partial k$, is close to zero, while, at the same frequency, the group velocity $v_\mathrm{g}\gg 0$ remains positive for $k>0$. One can observe a similar asymmetry at a frequency of \SI{12.0}{\giga\hertz}, where, only for $k<0$, there is a local minimum in the dispersion, and thus $v_\mathrm{g}\approx 0$ only for this propagation direction. The principle of the slow-wave-based magnonic diode now relies on the fact that the group velocity of spin waves is directly proportional to their attenuation length. In real samples, magnetic damping will induce a non-zero linewidth $\Gamma_{\nu}(k)\neq 0$ of the modes which leads to an exponential decay in propagation direction according to $\exp[ikz-z/L_\nu(k)])$, 
where $L_\nu$ is the attenuation length of the mode, related to the group velocity by $L_\nu=v_\mathrm{g}/\Gamma_\nu$. As a result, when exciting spin waves (\textit{e.g.} using a current-loop microwave antenna, see Fig.~\figref{fig:FIG7}{b}) at the center of the nanotube, in these particular cases, the modes for negative wave vectors will be strongly attenuated. To illustrate this, in Fig.~\figref{fig:FIG7}{c,d}, we show snapshots of the dynamical magnetization at the excitation frequencies \SI{10.7}{\giga\hertz} and \SI{12.0}{\giga\hertz} as cutouts in a plane which contains a radial direction and the axis of the nanotube sketched in Fig.~\figref{fig:FIG7}{b}. The snapshots were calculated  taking into account the corresponding lateral mode profiles $\bm{\eta}_\nu(k)$. For the calculations, we approximate the linewidths of the modes as $\Gamma_\nu(k) = \alpha_\mathrm{G}\omega_\nu(k)$ with a Gilbert-damping parameter of $\alpha_\mathrm{G}=0.01$. Furthermore, we neglect that the modes propagating in opposite directions have, in general, different dynamic susceptibilities. This, however, only leads to a scaling of the spin-wave profiles and not to a change in the attenuation lengths. As can be seen clearly in Figs.~\figref{fig:FIG7}{c,d}, the spin-wave propagation is nearly unidirectional, the waves propagating in $-z$ direction are strongly attenuated while, for the other direction, they propagate much further. Note that, here also, the radial profiles of the modes propagating in opposite directions are vastly different from each other due to the curvature-induced mode-profile asymmetry.

\subsection{Curvature effects on three-magnon splitting}

\begin{figure}[h!]
    \centering
    \includegraphics{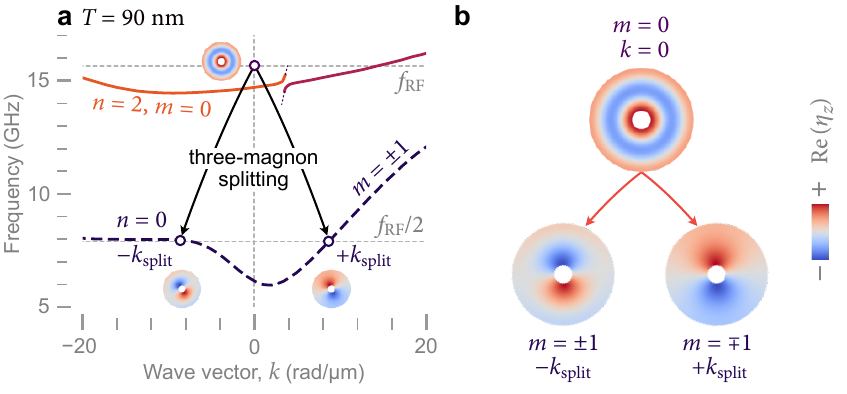}
    \caption{A possible channel for the three-magnon splitting is sketched in panel (a). The directly excited mode profile (at $k=0$) as well as the split mode profiles are included as insets and again shown magnified in panel (b).}
    \label{fig:FIG8}
\end{figure}
As an outlook on the nonlinear spin-wave dynamics in curved magnetic shells, here, we qualitatively discuss the lowest order of nonlinear spin-wave interaction -- three-magnon splitting -- where one primary spin wave, excited above a certain power threshold, splits into two secondary waves. In bulk ferromagnets and films three-magnon splitting obeys the conservation of energy and momentum,
\begin{equation}
    \begin{split}
        \omega_0 &= \omega_1 + \omega_2\\
        \bm{k}_0 &= \bm{k}_1 + \bm{k}_2
    \end{split}
\end{equation}
In confined magnetic objects the momentum conservation rule transforms into specific rules on the mode indices; also additional selection rules can appear depending on the geometry and magnetic state of a structure. For three-magnon splitting (as well as for the reversed process of three-wave confluence), selection rules are often quite strong. This happens because three-magnon interaction relies, typically, on the asymmetric part of magnetodipolar interaction, leading to strong restriction on the modes symmetry involved in the scattering process \cite{verbaTheoryThreemagnonInteraction2021, Etesamirad2021}. In particular, this type of spin-wave scattering has already been examined theoretically and experimentally in a system very closely related to our vortex-state nanotubes, namely in flat magnetic disks in the vortex state. \cite{schultheissExcitationWhisperingGallery, korberNonlocalStimulationThreemagnon2020, verbaTheoryThreemagnonInteraction2021} In this system it was found that radial modes ($m_0=0$) can split into two azimuthal modes with opposite azimuthal indices $m_1=-m_2$ (which constitutes the conservation of angular momentum) and different radial indices ($n_1\neq n_2$). This second selection rule of having different radial profiles follows from the symmetry of the magnetic system and leads for the secondary waves to be separated by a frequency gap ($\omega_{1,2} = \omega_0/2 \pm \Delta\omega$). Three-magnon splitting of a radial mode allowed, for the first time, for the  experimental observation of azimuthal modes with unprecedentedly large azimuthal index $m$ and, in the future, could be advantageous, for example, for physical reservoir computing. 

In vortex-state magnetic nanotubes, similar nonlinear dynamics is possible. Naturally, rotational symmetry manifests itself in the conservation of angular momentum or, in other words, of azimuthal index, $m_0 = m_1 + m_2$, the same as in the vortex-state dots. In addition, translational symmetry in $z$ direction results in the wave vector conservation rule $k_0 = k_1 + k_2$. To understand other possible selection rules one needs to look on the properties of the effective spin-wave tensor $\vu N_k$. Within the framework of vectorial Hamiltonian formalism for spin-wave dynamics \cite{Tyberkevych_ArXiv, verbaTheoryThreemagnonInteraction2021} the three-magnon part of the spin-wave Hamiltonian, which governs three-wave processes, is given by 
\begin{equation}
 \mathcal{H}^{(3)} = - \frac{\omega_M}{2S} \int |\bm{\eta}|^2 \bm{m}_0 \cdot \vu N_k \cdot \bm{\eta} d \bm{\rho} \,,
\end{equation}
where $S$ is the nanotube cross-section area. Since $\bm{\eta} \bot \bm{m}_0$, only off-diagonal components of the operator $\vu N_k$ could result in three-wave coupling. The exchange operator $\vu{N}_k^{(ex)} = -\lambda^2 (\nabla_{2\mathrm{D}}^2 -k^2) \mat I$ (written here in Cartesian coordinates) is modified only by an additive diagonal term compared to the dot case. Thus, the exchange part of three-magnon interaction efficiency possesses the same rules as in the case of vortex dots, in particular, in the case of splitting of radial modes it is nonzero only if the split modes are different in their radial profiles. As shown above, in nanotubes this difference appears due to the curvature-induced nonreciprocity for modes belonging to the same spin-wave branch (same radial index $n$) and is large, which results in relaxing of the selection rule $n_1 \neq n_2$ in comparison to the vortex-state disks.

The dipolar interaction, described by magnetostatic Green's function $\vu G_k$, $\vu N_k \cdot \bm{\eta} = \int \vu G_k (\bm{\rho}, \bm{\rho}') \cdot \bm{\eta}(\bm{\rho}') d\bm{\rho}'$, in contrast, is more modified compared to the disk case. In addition to the off-diagonal component $G_{\rho \varphi}$, the $G_{z\varphi}$ component becomes nonzero and, thus, relevant for three-magnon interaction (see full expression for Green function in the supplementary material). The selection rule of different radial profiles of split modes, which takes place in vortex dots, comes from the property $G_{k_1,m,\alpha\beta} = -G_{k_2, -m, \alpha\beta}$. For the splitting of spin-wave modes having $k = 0$ (thus, $k_1 = - k_2$) the $\rho \varphi$ component still posses this symmetry and, thus, only curvature-induced nonreciprocity of spin-wave profiles results in nonzero contribution to three-magnon interaction with secondary modes of the same branch. The component $G_{z\varphi}$ possesses different symmetry relations and the selection rule $n_1 \neq n_2$ is relaxed by the asymmetry of the dipolar interaction itself. The relative contribution of one or another off-diagonal term depends on the azimuthal and radial indices and wave numbers of the interacting modes, as well as on the geometry. For the case of splitting of propagating waves ($k_0 \neq 0$), a difference in the radial profiles of the secondary modes us not required for both dipolar contributions, which is quite natural since the modes becomes non-reciprocal by their wavenumbers.

Less strict rules for three-magnon splitting facilitates its applications in nanotubes. Since secondary waves can belong to the same, lowest spectral branch, directly excited (primary) wave could have a lower radial index and frequency compared to vortex-state disks and rings of the same radius, and, thus, are easier to be excited in an experiment. A viable application of three-magnon splitting is the excitation of high-$m$ azimuthal modes, which, to a certain extent, can bear topologically protected information.\cite{Yuan_APL2022_twisted_magnon} Another practically interesting case, which benefits from both the nonreciprocal mode hybridization and enriched possibilities for three-magnon scattering is sketched in Fig.~\ref{fig:FIG8}.  As an example, in a tube with $T=\SI{90}{\nano\meter}$ mantel thickness, we schematically show the three-magnon splitting of a radial mode which is homogeneous along the nanotube axis ($m=k=0$), as seen in Fig.~\figref{fig:FIG8}{a}. The mode is excited at a microwave frequency $f_\mathrm{RF}$ which is a little bit larger than the mode resonance frequency. This is done in order to parametrically excite desired secondary modes at $f_\mathrm{RF}/2$. As known from flat disks in the vortex state, this slightly off-resonant excitation of the primary mode can still lead to an efficient triggering of three-magnon splitting. By choosing the direct excitation frequency $f_\mathrm{RF}$ it is possible to have for one of the potential secondary waves, like in the previous section, vanishing group velocity (see Fig.~\figref{fig:FIG7}{a}). \textit{Nota bene}, many other splitting channels (e.g., modes with different azimuthal numbers) for this particular excitation frequency might be allowed and the splitting channel discussed here is not necessarily the one with the lowest power threshold, \textit{i.e.} not necessarily the channel which is triggered \textit{first} when increasing the excitation power. Determining the channel with the lowest threshold, however, would require a quantitative analysis of the three-magnon-splitting coefficients (analogous to, \textit{e.g.} Ref.~\citenum{verbaTheoryThreemagnonInteraction2021}) and would go far beyond the scope of this work.

\section{Conclusions}

In conclusion, we have investigated the spin-wave dispersion for thick shells using the specific example of a vortex state magnetic nanotubes by micromagnetic simulations. We show that the dispersion changes quite drastically when transitioning to thick shells. The simulations confirm the existence of perpendicular standing spin waves similar to those known from flat thick films, and the curvature-induced dispersion asymmetry. However, with careful analysis of the mode profiles of the different spin-wave branches, we reveal an additional asymmetry in the mode profiles along the thickness of the shell, leading non-reciprocity of all spin-wave mode profiles, including higher-order ones such as PSSWs. The origin of the mode profile asymmetry can be explained within the concept of magnetic pseudo-charges, namely it is caused by the inhomogeneity of the extrinsic geometrical charge along the thickness of the shell. The consequence of this asymmetry is the nonreciprocal hybridization of modes with the same azimuthal mode number, which in return leads to asymmetric plateaus in the dispersion. The existence of these plateaus can be exploited, for example, to facilitate unidirectional spin-wave transport. In addition we show, that the mode profile asymmetry will also have consequences for the nonlinear spin-wave dynamics. For instance, will lead to the opening of additional splitting channels for the three-magnon scattering and allows for the splitting into modes propagating effectively in one direction only. 

We believe that these results emphasise that thick curved magnetic shells are not only promising spin-wave conduits for magnonic applications, because of their reach spin-wave dispersion, but due to the consequences of the curvature-induced asymmetric mode profiles on the nonlinear spin-wave dynamics, their application in reservoir computing -- the particular case of a thick nanotube -- can also be important.

\textit{Authors note:} In the final preparation phase of this paper, a similar study has been published by Gallardo \textit{et al.}\cite{gallardoHighSpinwaveAsymmetry2022} who also reported on the spin-wave dispersion in thick vortex-state nanotubes. Their results also highlight the emergence of higher-order radial modes but, subsequently, focused more on the increased frequency non-reciprocity. The two studies, namely the one in the mentioned publication and that of the current paper were carried out independently.

\section*{Supplementary materials}

See supplementary material, which includes Refs. \citenum{korberNumericalReverseEngineering2021,Guslienko_JAP2000,Guslienko_JMMM2011} and \citenum{Arfken_Book}, for an extended Fig.~\ref{fig:FIG4} with more thicknesses, for the thickness dependence of the spin-wave dispersion in a nanotube according to thin-shell theory, for a movie showing the continuous transition of the normal-mode dispersion, for a demonstration why the standard perturbation approach to calculate the dipole-exchange spectrum fails for thick tubes due to the strong dipolar influence on the mode profiles, and for expressions for the magnetostatic Green's function in thick nanotubes.

\section*{Acknowledgements}

The authors are thankful to Istv\'an K\'ezsm\'arki for fruitful discussions. Moreover, we truly appreciate Gabriella Gon\c{c}alles for the drawings of the hands in Fig.~\ref{fig:FIG1}. Financial support by the Deutsche Forschungsgemeinschaft within the program KA 5069/1-1 and KA 5069/3-1 are gratefully acknowledged. R. V. and V. K. acknowledge support by the National Academy of Sciences of Ukraine, projects \# 0122U001845 and \# 0120U100855, respectively.

%\bibliography{references.bib}

\begin{thebibliography}{70}%
\makeatletter
\providecommand \@ifxundefined [1]{%
 \@ifx{#1\undefined}
}%
\providecommand \@ifnum [1]{%
 \ifnum #1\expandafter \@firstoftwo
 \else \expandafter \@secondoftwo
 \fi
}%
\providecommand \@ifx [1]{%
 \ifx #1\expandafter \@firstoftwo
 \else \expandafter \@secondoftwo
 \fi
}%
\providecommand \natexlab [1]{#1}%
\providecommand \enquote  [1]{``#1''}%
\providecommand \bibnamefont  [1]{#1}%
\providecommand \bibfnamefont [1]{#1}%
\providecommand \citenamefont [1]{#1}%
\providecommand \href@noop [0]{\@secondoftwo}%
\providecommand \href [0]{\begingroup \@sanitize@url \@href}%
\providecommand \@href[1]{\@@startlink{#1}\@@href}%
\providecommand \@@href[1]{\endgroup#1\@@endlink}%
\providecommand \@sanitize@url [0]{\catcode `\\12\catcode `\$12\catcode
  `\&12\catcode `\#12\catcode `\^12\catcode `\_12\catcode `\%12\relax}%
\providecommand \@@startlink[1]{}%
\providecommand \@@endlink[0]{}%
\providecommand \url  [0]{\begingroup\@sanitize@url \@url }%
\providecommand \@url [1]{\endgroup\@href {#1}{\urlprefix }}%
\providecommand \urlprefix  [0]{URL }%
\providecommand \Eprint [0]{\href }%
\providecommand \doibase [0]{https://doi.org/}%
\providecommand \selectlanguage [0]{\@gobble}%
\providecommand \bibinfo  [0]{\@secondoftwo}%
\providecommand \bibfield  [0]{\@secondoftwo}%
\providecommand \translation [1]{[#1]}%
\providecommand \BibitemOpen [0]{}%
\providecommand \bibitemStop [0]{}%
\providecommand \bibitemNoStop [0]{.\EOS\space}%
\providecommand \EOS [0]{\spacefactor3000\relax}%
\providecommand \BibitemShut  [1]{\csname bibitem#1\endcsname}%
\let\auto@bib@innerbib\@empty
%</preamble>
\bibitem [{\citenamefont {Gaididei}, \citenamefont {Kravchuk},\ and\
  \citenamefont {Sheka}(2014)}]{gaidideiCurvatureEffectsThin2014a}%
  \BibitemOpen
  \bibfield  {author} {\bibinfo {author} {\bibfnamefont {Y.}~\bibnamefont
  {Gaididei}}, \bibinfo {author} {\bibfnamefont {V.~P.}\ \bibnamefont
  {Kravchuk}},\ and\ \bibinfo {author} {\bibfnamefont {D.~D.}\ \bibnamefont
  {Sheka}},\ }\bibfield  {title} {\enquote {\bibinfo {title} {Curvature
  {{Effects}} in {{Thin Magnetic Shells}}},}\ }\href
  {https://doi.org/10.1103/PhysRevLett.112.257203} {\bibfield  {journal}
  {\bibinfo  {journal} {Physical Review Letters}\ }\textbf {\bibinfo {volume}
  {112}},\ \bibinfo {pages} {257203} (\bibinfo {year} {2014})}\BibitemShut
  {NoStop}%
\bibitem [{\citenamefont {Sheka}, \citenamefont {Kravchuk},\ and\ \citenamefont
  {Gaididei}(2015)}]{shekaCurvatureEffectsStatics2015}%
  \BibitemOpen
  \bibfield  {author} {\bibinfo {author} {\bibfnamefont {D.~D.}\ \bibnamefont
  {Sheka}}, \bibinfo {author} {\bibfnamefont {V.~P.}\ \bibnamefont
  {Kravchuk}},\ and\ \bibinfo {author} {\bibfnamefont {Y.}~\bibnamefont
  {Gaididei}},\ }\bibfield  {title} {\enquote {\bibinfo {title} {Curvature
  effects in statics and dynamics of low dimensional magnets},}\ }\href
  {https://doi.org/10.1088/1751-8113/48/12/125202} {\bibfield  {journal}
  {\bibinfo  {journal} {Journal of Physics A: Mathematical and Theoretical}\
  }\textbf {\bibinfo {volume} {48}},\ \bibinfo {pages} {125202} (\bibinfo
  {year} {2015})}\BibitemShut {NoStop}%
\bibitem [{\citenamefont {Sheka}\ \emph {et~al.}(2015)\citenamefont {Sheka},
  \citenamefont {Kravchuk}, \citenamefont {Yershov},\ and\ \citenamefont
  {Gaididei}}]{shekaTorsioninducedEffectsMagnetic2015a}%
  \BibitemOpen
  \bibfield  {author} {\bibinfo {author} {\bibfnamefont {D.~D.}\ \bibnamefont
  {Sheka}}, \bibinfo {author} {\bibfnamefont {V.~P.}\ \bibnamefont {Kravchuk}},
  \bibinfo {author} {\bibfnamefont {K.~V.}\ \bibnamefont {Yershov}},\ and\
  \bibinfo {author} {\bibfnamefont {Y.}~\bibnamefont {Gaididei}},\ }\bibfield
  {title} {\enquote {\bibinfo {title} {Torsion-induced effects in magnetic
  nanowires},}\ }\href {https://doi.org/10.1103/PhysRevB.92.054417} {\bibfield
  {journal} {\bibinfo  {journal} {Physical Review B}\ }\textbf {\bibinfo
  {volume} {92}},\ \bibinfo {pages} {054417} (\bibinfo {year}
  {2015})}\BibitemShut {NoStop}%
\bibitem [{\citenamefont {Korniienko}\ \emph {et~al.}(2019)\citenamefont
  {Korniienko}, \citenamefont {Kravchuk}, \citenamefont {Pylypovskyi},
  \citenamefont {Sheka}, \citenamefont {{van den Brink}},\ and\ \citenamefont
  {Gaididei}}]{korniienkoCurvatureInducedMagnonic2019}%
  \BibitemOpen
  \bibfield  {author} {\bibinfo {author} {\bibfnamefont {A.}~\bibnamefont
  {Korniienko}}, \bibinfo {author} {\bibfnamefont {V.}~\bibnamefont
  {Kravchuk}}, \bibinfo {author} {\bibfnamefont {O.}~\bibnamefont
  {Pylypovskyi}}, \bibinfo {author} {\bibfnamefont {D.}~\bibnamefont {Sheka}},
  \bibinfo {author} {\bibfnamefont {J.}~\bibnamefont {{van den Brink}}},\ and\
  \bibinfo {author} {\bibfnamefont {Y.}~\bibnamefont {Gaididei}},\ }\bibfield
  {title} {\enquote {\bibinfo {title} {Curvature induced magnonic crystal in
  nanowires},}\ }\href {https://doi.org/10.21468/scipostphys.7.3.035}
  {\bibfield  {journal} {\bibinfo  {journal} {SciPost Physics}\ }\textbf
  {\bibinfo {volume} {7}},\ \bibinfo {pages} {1--19} (\bibinfo {year}
  {2019})}\BibitemShut {NoStop}%
\bibitem [{\citenamefont {Bord{\'a}cs}\ \emph {et~al.}(2012)\citenamefont
  {Bord{\'a}cs}, \citenamefont {K{\'e}zsm{\'a}rki}, \citenamefont {Szaller},
  \citenamefont {Demk{\'o}}, \citenamefont {Kida}, \citenamefont {Murakawa},
  \citenamefont {Onose}, \citenamefont {Shimano}, \citenamefont {R{\~o}{\~o}m},
  \citenamefont {Nagel}, \citenamefont {Miyahara}, \citenamefont {Furukawa},\
  and\ \citenamefont {Tokura}}]{bordacsChiralityMatterShows2012a}%
  \BibitemOpen
  \bibfield  {author} {\bibinfo {author} {\bibfnamefont {S.}~\bibnamefont
  {Bord{\'a}cs}}, \bibinfo {author} {\bibfnamefont {I.}~\bibnamefont
  {K{\'e}zsm{\'a}rki}}, \bibinfo {author} {\bibfnamefont {D.}~\bibnamefont
  {Szaller}}, \bibinfo {author} {\bibfnamefont {L.}~\bibnamefont {Demk{\'o}}},
  \bibinfo {author} {\bibfnamefont {N.}~\bibnamefont {Kida}}, \bibinfo {author}
  {\bibfnamefont {H.}~\bibnamefont {Murakawa}}, \bibinfo {author}
  {\bibfnamefont {Y.}~\bibnamefont {Onose}}, \bibinfo {author} {\bibfnamefont
  {R.}~\bibnamefont {Shimano}}, \bibinfo {author} {\bibfnamefont
  {T.}~\bibnamefont {R{\~o}{\~o}m}}, \bibinfo {author} {\bibfnamefont
  {U.}~\bibnamefont {Nagel}}, \bibinfo {author} {\bibfnamefont
  {S.}~\bibnamefont {Miyahara}}, \bibinfo {author} {\bibfnamefont
  {N.}~\bibnamefont {Furukawa}},\ and\ \bibinfo {author} {\bibfnamefont
  {Y.}~\bibnamefont {Tokura}},\ }\bibfield  {title} {\enquote {\bibinfo {title}
  {Chirality of matter shows up via spin excitations},}\ }\href
  {https://doi.org/10.1038/nphys2387} {\bibfield  {journal} {\bibinfo
  {journal} {Nature Physics}\ }\textbf {\bibinfo {volume} {8}},\ \bibinfo
  {pages} {734--738} (\bibinfo {year} {2012})}\BibitemShut {NoStop}%
\bibitem [{\citenamefont
  {Hertel}(2013)}]{hertelCurvatureInducedMagnetochirality2013}%
  \BibitemOpen
  \bibfield  {author} {\bibinfo {author} {\bibfnamefont {R.}~\bibnamefont
  {Hertel}},\ }\bibfield  {title} {\enquote {\bibinfo {title}
  {Curvature\textendash induced magnetochirality},}\ }\href
  {https://doi.org/10.1142/s2010324713400092} {\bibfield  {journal} {\bibinfo
  {journal} {SPIN}\ }\textbf {\bibinfo {volume} {03}},\ \bibinfo {pages}
  {1340009} (\bibinfo {year} {2013})}\BibitemShut {NoStop}%
\bibitem [{\citenamefont {K{\'e}zsm{\'a}rki}\ \emph
  {et~al.}(2014{\natexlab{a}})\citenamefont {K{\'e}zsm{\'a}rki}, \citenamefont
  {Szaller}, \citenamefont {Bord{\'a}cs}, \citenamefont {Kocsis}, \citenamefont
  {Tokunaga}, \citenamefont {Taguchi}, \citenamefont {Murakawa}, \citenamefont
  {Tokura}, \citenamefont {Engelkamp}, \citenamefont {R{\~o}{\~o}m},\ and\
  \citenamefont {Nagel}}]{kezsmarkiOnewayTransparencyFourcoloured2014a}%
  \BibitemOpen
  \bibfield  {author} {\bibinfo {author} {\bibfnamefont {I.}~\bibnamefont
  {K{\'e}zsm{\'a}rki}}, \bibinfo {author} {\bibfnamefont {D.}~\bibnamefont
  {Szaller}}, \bibinfo {author} {\bibfnamefont {S.}~\bibnamefont
  {Bord{\'a}cs}}, \bibinfo {author} {\bibfnamefont {V.}~\bibnamefont {Kocsis}},
  \bibinfo {author} {\bibfnamefont {Y.}~\bibnamefont {Tokunaga}}, \bibinfo
  {author} {\bibfnamefont {Y.}~\bibnamefont {Taguchi}}, \bibinfo {author}
  {\bibfnamefont {H.}~\bibnamefont {Murakawa}}, \bibinfo {author}
  {\bibfnamefont {Y.}~\bibnamefont {Tokura}}, \bibinfo {author} {\bibfnamefont
  {H.}~\bibnamefont {Engelkamp}}, \bibinfo {author} {\bibfnamefont
  {T.}~\bibnamefont {R{\~o}{\~o}m}},\ and\ \bibinfo {author} {\bibfnamefont
  {U.}~\bibnamefont {Nagel}},\ }\bibfield  {title} {\enquote {\bibinfo {title}
  {One-way transparency of four-coloured spin-wave excitations in multiferroic
  materials},}\ }\href {https://doi.org/10.1038/ncomms4203} {\bibfield
  {journal} {\bibinfo  {journal} {Nature Communications}\ }\textbf {\bibinfo
  {volume} {5}},\ \bibinfo {pages} {3203} (\bibinfo {year}
  {2014}{\natexlab{a}})}\BibitemShut {NoStop}%
\bibitem [{\citenamefont {Sheka}\ \emph {et~al.}(2020)\citenamefont {Sheka},
  \citenamefont {Pylypovskyi}, \citenamefont {Landeros}, \citenamefont
  {Gaididei}, \citenamefont {K{\'a}kay},\ and\ \citenamefont
  {Makarov}}]{shekaNonlocalChiralSymmetry2020}%
  \BibitemOpen
  \bibfield  {author} {\bibinfo {author} {\bibfnamefont {D.~D.}\ \bibnamefont
  {Sheka}}, \bibinfo {author} {\bibfnamefont {O.~V.}\ \bibnamefont
  {Pylypovskyi}}, \bibinfo {author} {\bibfnamefont {P.}~\bibnamefont
  {Landeros}}, \bibinfo {author} {\bibfnamefont {Y.}~\bibnamefont {Gaididei}},
  \bibinfo {author} {\bibfnamefont {A.}~\bibnamefont {K{\'a}kay}},\ and\
  \bibinfo {author} {\bibfnamefont {D.}~\bibnamefont {Makarov}},\ }\bibfield
  {title} {\enquote {\bibinfo {title} {Nonlocal chiral symmetry breaking in
  curvilinear magnetic shells},}\ }\href
  {https://doi.org/10.1038/s42005-020-0387-2} {\bibfield  {journal} {\bibinfo
  {journal} {Communications Physics}\ }\textbf {\bibinfo {volume} {3}},\
  \bibinfo {pages} {128} (\bibinfo {year} {2020})}\BibitemShut {NoStop}%
\bibitem [{\citenamefont {Kravchuk}\ \emph {et~al.}(2018)\citenamefont
  {Kravchuk}, \citenamefont {Sheka}, \citenamefont {K{\'a}kay}, \citenamefont
  {Volkov}, \citenamefont {R{\"o}{\ss}ler}, \citenamefont {{van den Brink}},
  \citenamefont {Makarov},\ and\ \citenamefont
  {Gaididei}}]{kravchukMultipletSkyrmionStates2018}%
  \BibitemOpen
  \bibfield  {author} {\bibinfo {author} {\bibfnamefont {V.~P.}\ \bibnamefont
  {Kravchuk}}, \bibinfo {author} {\bibfnamefont {D.~D.}\ \bibnamefont {Sheka}},
  \bibinfo {author} {\bibfnamefont {A.}~\bibnamefont {K{\'a}kay}}, \bibinfo
  {author} {\bibfnamefont {O.~M.}\ \bibnamefont {Volkov}}, \bibinfo {author}
  {\bibfnamefont {U.~K.}\ \bibnamefont {R{\"o}{\ss}ler}}, \bibinfo {author}
  {\bibfnamefont {J.}~\bibnamefont {{van den Brink}}}, \bibinfo {author}
  {\bibfnamefont {D.}~\bibnamefont {Makarov}},\ and\ \bibinfo {author}
  {\bibfnamefont {Y.}~\bibnamefont {Gaididei}},\ }\bibfield  {title} {\enquote
  {\bibinfo {title} {Multiplet of {{Skyrmion States}} on a {{Curvilinear
  Defect}}: {{Reconfigurable Skyrmion Lattices}}},}\ }\href
  {https://doi.org/10.1103/PhysRevLett.120.067201} {\bibfield  {journal}
  {\bibinfo  {journal} {Phys. Rev. Lett.}\ }\textbf {\bibinfo {volume} {120}},\
  \bibinfo {pages} {67201} (\bibinfo {year} {2018})}\BibitemShut {NoStop}%
\bibitem [{\citenamefont {El{\'i}as}, \citenamefont {{Vidal-Silva}},\ and\
  \citenamefont {{Carvalho-Santos}}(2019)}]{eliasWindingNumberSelection2019}%
  \BibitemOpen
  \bibfield  {author} {\bibinfo {author} {\bibfnamefont {R.~G.}\ \bibnamefont
  {El{\'i}as}}, \bibinfo {author} {\bibfnamefont {N.}~\bibnamefont
  {{Vidal-Silva}}},\ and\ \bibinfo {author} {\bibfnamefont {V.~L.}\
  \bibnamefont {{Carvalho-Santos}}},\ }\bibfield  {title} {\enquote {\bibinfo
  {title} {Winding number selection on merons by {{Gaussian}} curvature's
  sign},}\ }\href {https://doi.org/10.1038/s41598-019-50395-7} {\bibfield
  {journal} {\bibinfo  {journal} {Scientific Reports}\ }\textbf {\bibinfo
  {volume} {9}},\ \bibinfo {pages} {14309} (\bibinfo {year}
  {2019})}\BibitemShut {NoStop}%
\bibitem [{\citenamefont {Lewis}\ \emph {et~al.}(2009)\citenamefont {Lewis},
  \citenamefont {Petit}, \citenamefont {Thevenard}, \citenamefont {Jausovec},
  \citenamefont {O'Brien}, \citenamefont {Read},\ and\ \citenamefont
  {Cowburn}}]{lewisMagneticDomainWall2009}%
  \BibitemOpen
  \bibfield  {author} {\bibinfo {author} {\bibfnamefont {E.~R.}\ \bibnamefont
  {Lewis}}, \bibinfo {author} {\bibfnamefont {D.}~\bibnamefont {Petit}},
  \bibinfo {author} {\bibfnamefont {L.}~\bibnamefont {Thevenard}}, \bibinfo
  {author} {\bibfnamefont {A.~V.}\ \bibnamefont {Jausovec}}, \bibinfo {author}
  {\bibfnamefont {L.}~\bibnamefont {O'Brien}}, \bibinfo {author} {\bibfnamefont
  {D.~E.}\ \bibnamefont {Read}},\ and\ \bibinfo {author} {\bibfnamefont
  {R.~P.}\ \bibnamefont {Cowburn}},\ }\bibfield  {title} {\enquote {\bibinfo
  {title} {Magnetic domain wall pinning by a curved conduit},}\ }\href
  {https://doi.org/10.1063/1.3246154} {\bibfield  {journal} {\bibinfo
  {journal} {Appl. Phys. Lett.}\ }\textbf {\bibinfo {volume} {95}},\ \bibinfo
  {pages} {152505} (\bibinfo {year} {2009})}\BibitemShut {NoStop}%
\bibitem [{\citenamefont {Yershov}\ \emph {et~al.}(2015)\citenamefont
  {Yershov}, \citenamefont {Kravchuk}, \citenamefont {Sheka},\ and\
  \citenamefont {Gaididei}}]{yershovCurvatureInducedDomain2015}%
  \BibitemOpen
  \bibfield  {author} {\bibinfo {author} {\bibfnamefont {K.~V.}\ \bibnamefont
  {Yershov}}, \bibinfo {author} {\bibfnamefont {V.~P.}\ \bibnamefont
  {Kravchuk}}, \bibinfo {author} {\bibfnamefont {D.~D.}\ \bibnamefont
  {Sheka}},\ and\ \bibinfo {author} {\bibfnamefont {Y.}~\bibnamefont
  {Gaididei}},\ }\bibfield  {title} {\enquote {\bibinfo {title} {Curvature
  induced domain wall pinning},}\ }\href@noop {} {\bibfield  {journal}
  {\bibinfo  {journal} {ArXiv e-prints}\ } (\bibinfo {year}
  {2015})}\BibitemShut {NoStop}%
\bibitem [{\citenamefont {Volkov}\ \emph {et~al.}(2019)\citenamefont {Volkov},
  \citenamefont {K{\'a}kay}, \citenamefont {Kronast}, \citenamefont
  {M{\"o}nch}, \citenamefont {Mawass}, \citenamefont {Fassbender},\ and\
  \citenamefont {Makarov}}]{volkovExperimentalObservationExchangeDriven2019}%
  \BibitemOpen
  \bibfield  {author} {\bibinfo {author} {\bibfnamefont {O.~M.}\ \bibnamefont
  {Volkov}}, \bibinfo {author} {\bibfnamefont {A.}~\bibnamefont {K{\'a}kay}},
  \bibinfo {author} {\bibfnamefont {F.}~\bibnamefont {Kronast}}, \bibinfo
  {author} {\bibfnamefont {I.}~\bibnamefont {M{\"o}nch}}, \bibinfo {author}
  {\bibfnamefont {M.~A.}\ \bibnamefont {Mawass}}, \bibinfo {author}
  {\bibfnamefont {J.}~\bibnamefont {Fassbender}},\ and\ \bibinfo {author}
  {\bibfnamefont {D.}~\bibnamefont {Makarov}},\ }\bibfield  {title} {\enquote
  {\bibinfo {title} {Experimental {{Observation}} of {{Exchange-Driven Chiral
  Effects}} in {{Curvilinear Magnetism}}},}\ }\href
  {https://doi.org/10.1103/PhysRevLett.123.077201} {\bibfield  {journal}
  {\bibinfo  {journal} {Physical Review Letters}\ }\textbf {\bibinfo {volume}
  {123}},\ \bibinfo {pages} {77201} (\bibinfo {year} {2019})}\BibitemShut
  {NoStop}%
\bibitem [{\citenamefont {Gaididei}\ \emph {et~al.}(2018)\citenamefont
  {Gaididei}, \citenamefont {Kravchuk}, \citenamefont {Mertens}, \citenamefont
  {Pylypovskyi}, \citenamefont {Saxena}, \citenamefont {Sheka},\ and\
  \citenamefont {Volkov}}]{gaidideiLocalizationMagnonModes2018}%
  \BibitemOpen
  \bibfield  {author} {\bibinfo {author} {\bibfnamefont {Y.}~\bibnamefont
  {Gaididei}}, \bibinfo {author} {\bibfnamefont {V.~P.}\ \bibnamefont
  {Kravchuk}}, \bibinfo {author} {\bibfnamefont {F.~G.}\ \bibnamefont
  {Mertens}}, \bibinfo {author} {\bibfnamefont {O.~V.}\ \bibnamefont
  {Pylypovskyi}}, \bibinfo {author} {\bibfnamefont {A.}~\bibnamefont {Saxena}},
  \bibinfo {author} {\bibfnamefont {D.~D.}\ \bibnamefont {Sheka}},\ and\
  \bibinfo {author} {\bibfnamefont {O.~M.}\ \bibnamefont {Volkov}},\ }\bibfield
   {title} {\enquote {\bibinfo {title} {Localization of magnon modes in a
  curved magnetic nanowire},}\ }\href {https://doi.org/10.1063/1.5041428}
  {\bibfield  {journal} {\bibinfo  {journal} {Low Temperature Physics}\
  }\textbf {\bibinfo {volume} {44}},\ \bibinfo {pages} {814--823} (\bibinfo
  {year} {2018})}\BibitemShut {NoStop}%
\bibitem [{\citenamefont {Ot{\'a}lora}\ \emph {et~al.}(2016)\citenamefont
  {Ot{\'a}lora}, \citenamefont {Yan}, \citenamefont {Schultheiss},
  \citenamefont {Hertel},\ and\ \citenamefont
  {K{\'a}kay}}]{otaloraCurvatureInducedAsymmetricSpinWave2016}%
  \BibitemOpen
  \bibfield  {author} {\bibinfo {author} {\bibfnamefont {J.~A.}\ \bibnamefont
  {Ot{\'a}lora}}, \bibinfo {author} {\bibfnamefont {M.}~\bibnamefont {Yan}},
  \bibinfo {author} {\bibfnamefont {H.}~\bibnamefont {Schultheiss}}, \bibinfo
  {author} {\bibfnamefont {R.}~\bibnamefont {Hertel}},\ and\ \bibinfo {author}
  {\bibfnamefont {A.}~\bibnamefont {K{\'a}kay}},\ }\bibfield  {title} {\enquote
  {\bibinfo {title} {Curvature-{{Induced Asymmetric Spin-Wave Dispersion}}},}\
  }\href {https://doi.org/10.1103/PhysRevLett.117.227203} {\bibfield  {journal}
  {\bibinfo  {journal} {Physical Review Letters}\ }\textbf {\bibinfo {volume}
  {117}},\ \bibinfo {pages} {227203} (\bibinfo {year} {2016})}\BibitemShut
  {NoStop}%
\bibitem [{\citenamefont {Ot{\'a}lora}\ \emph {et~al.}(2017)\citenamefont
  {Ot{\'a}lora}, \citenamefont {Yan}, \citenamefont {Schultheiss},
  \citenamefont {Hertel},\ and\ \citenamefont
  {K{\'a}kay}}]{otaloraAsymmetricSpinwaveDispersion2017}%
  \BibitemOpen
  \bibfield  {author} {\bibinfo {author} {\bibfnamefont {J.~A.}\ \bibnamefont
  {Ot{\'a}lora}}, \bibinfo {author} {\bibfnamefont {M.}~\bibnamefont {Yan}},
  \bibinfo {author} {\bibfnamefont {H.}~\bibnamefont {Schultheiss}}, \bibinfo
  {author} {\bibfnamefont {R.}~\bibnamefont {Hertel}},\ and\ \bibinfo {author}
  {\bibfnamefont {A.}~\bibnamefont {K{\'a}kay}},\ }\bibfield  {title} {\enquote
  {\bibinfo {title} {Asymmetric spin-wave dispersion in ferromagnetic nanotubes
  induced by surface curvature},}\ }\href
  {https://doi.org/10.1103/PhysRevB.95.184415} {\bibfield  {journal} {\bibinfo
  {journal} {PHYSICAL REVIEW B}\ }\textbf {\bibinfo {volume} {95}},\ \bibinfo
  {pages} {184415} (\bibinfo {year} {2017})}\BibitemShut {NoStop}%
\bibitem [{\citenamefont {{Cort{\'e}s-Ortu{\~n}o}}\ and\ \citenamefont
  {Landeros}(2013)}]{cortes-ortunoInfluenceDzyaloshinskiiMoriya2013}%
  \BibitemOpen
  \bibfield  {author} {\bibinfo {author} {\bibfnamefont {D.}~\bibnamefont
  {{Cort{\'e}s-Ortu{\~n}o}}}\ and\ \bibinfo {author} {\bibfnamefont
  {P.}~\bibnamefont {Landeros}},\ }\bibfield  {title} {\enquote {\bibinfo
  {title} {Influence of the {{Dzyaloshinskii}}\textendash{{Moriya}} interaction
  on the spin-wave spectra of thin films},}\ }\href
  {https://doi.org/10.1088/0953-8984/25/15/156001} {\bibfield  {journal}
  {\bibinfo  {journal} {Journal of Physics: Condensed Matter}\ }\textbf
  {\bibinfo {volume} {25}},\ \bibinfo {pages} {156001} (\bibinfo {year}
  {2013})}\BibitemShut {NoStop}%
\bibitem [{\citenamefont {Moon}\ \emph {et~al.}(2013)\citenamefont {Moon},
  \citenamefont {Seo}, \citenamefont {Lee}, \citenamefont {Kim}, \citenamefont
  {Ryu}, \citenamefont {Lee}, \citenamefont {McMichael},\ and\ \citenamefont
  {Stiles}}]{moonSpinwavePropagationPresence2013}%
  \BibitemOpen
  \bibfield  {author} {\bibinfo {author} {\bibfnamefont {J.~H.}\ \bibnamefont
  {Moon}}, \bibinfo {author} {\bibfnamefont {S.~M.}\ \bibnamefont {Seo}},
  \bibinfo {author} {\bibfnamefont {K.~J.}\ \bibnamefont {Lee}}, \bibinfo
  {author} {\bibfnamefont {K.~W.}\ \bibnamefont {Kim}}, \bibinfo {author}
  {\bibfnamefont {J.}~\bibnamefont {Ryu}}, \bibinfo {author} {\bibfnamefont
  {H.~W.}\ \bibnamefont {Lee}}, \bibinfo {author} {\bibfnamefont {R.~D.}\
  \bibnamefont {McMichael}},\ and\ \bibinfo {author} {\bibfnamefont {M.~D.}\
  \bibnamefont {Stiles}},\ }\bibfield  {title} {\enquote {\bibinfo {title}
  {Spin-wave propagation in the presence of interfacial
  {{Dzyaloshinskii-Moriya}} interaction},}\ }\href
  {https://doi.org/10.1103/PhysRevB.88.184404} {\bibfield  {journal} {\bibinfo
  {journal} {Physical Review B - Condensed Matter and Materials Physics}\
  }\textbf {\bibinfo {volume} {88}},\ \bibinfo {pages} {184404} (\bibinfo
  {year} {2013})}\BibitemShut {NoStop}%
\bibitem [{\citenamefont {Mika}\ and\ \citenamefont
  {Gr{\"u}nberg}(1985)}]{mikaDipolarSpinwaveModes1985}%
  \BibitemOpen
  \bibfield  {author} {\bibinfo {author} {\bibfnamefont {K.}~\bibnamefont
  {Mika}}\ and\ \bibinfo {author} {\bibfnamefont {P.}~\bibnamefont
  {Gr{\"u}nberg}},\ }\bibfield  {title} {\enquote {\bibinfo {title} {Dipolar
  spin-wave modes of a ferromagnetic multilayer with alternating directions of
  magnetization},}\ }\href {https://doi.org/10.1103/PhysRevB.31.4465}
  {\bibfield  {journal} {\bibinfo  {journal} {Physical Review B}\ }\textbf
  {\bibinfo {volume} {31}},\ \bibinfo {pages} {4465--4471} (\bibinfo {year}
  {1985})}\BibitemShut {NoStop}%
\bibitem [{\citenamefont {Gr{\"u}nberg}(1985)}]{grunbergWaysModifySpin1985}%
  \BibitemOpen
  \bibfield  {author} {\bibinfo {author} {\bibfnamefont {P.}~\bibnamefont
  {Gr{\"u}nberg}},\ }\bibfield  {title} {\enquote {\bibinfo {title} {Some ways
  to modify the spin-wave mode spectra of magnetic multilayers (invited)},}\
  }\href {https://doi.org/10.1063/1.334985} {\bibfield  {journal} {\bibinfo
  {journal} {Journal of Applied Physics}\ }\textbf {\bibinfo {volume} {57}},\
  \bibinfo {pages} {3673--3677} (\bibinfo {year} {1985})}\BibitemShut {NoStop}%
\bibitem [{\citenamefont {Gr{\"u}nberg}\ \emph {et~al.}(1986)\citenamefont
  {Gr{\"u}nberg}, \citenamefont {Schreiber}, \citenamefont {Pang},
  \citenamefont {Brodsky},\ and\ \citenamefont
  {Sowers}}]{grunbergLayeredMagneticStructures1986}%
  \BibitemOpen
  \bibfield  {author} {\bibinfo {author} {\bibfnamefont {P.}~\bibnamefont
  {Gr{\"u}nberg}}, \bibinfo {author} {\bibfnamefont {R.}~\bibnamefont
  {Schreiber}}, \bibinfo {author} {\bibfnamefont {Y.}~\bibnamefont {Pang}},
  \bibinfo {author} {\bibfnamefont {M.~B.}\ \bibnamefont {Brodsky}},\ and\
  \bibinfo {author} {\bibfnamefont {H.}~\bibnamefont {Sowers}},\ }\bibfield
  {title} {\enquote {\bibinfo {title} {Layered {{Magnetic Structures}}:
  {{Evidence}} for {{Antiferromagnetic Coupling}} of {{Fe Layers}} across {{Cr
  Interlayers}}},}\ }\href {https://doi.org/10.1103/PhysRevLett.57.2442}
  {\bibfield  {journal} {\bibinfo  {journal} {Physical Review Letters}\
  }\textbf {\bibinfo {volume} {57}},\ \bibinfo {pages} {2442--2445} (\bibinfo
  {year} {1986})}\BibitemShut {NoStop}%
\bibitem [{\citenamefont {Gladii}\ \emph {et~al.}(2016)\citenamefont {Gladii},
  \citenamefont {Haidar}, \citenamefont {Henry}, \citenamefont {Kostylev},\
  and\ \citenamefont {Bailleul}}]{gladiiFrequencyNonreciprocitySurface2016}%
  \BibitemOpen
  \bibfield  {author} {\bibinfo {author} {\bibfnamefont {O.}~\bibnamefont
  {Gladii}}, \bibinfo {author} {\bibfnamefont {M.}~\bibnamefont {Haidar}},
  \bibinfo {author} {\bibfnamefont {Y.}~\bibnamefont {Henry}}, \bibinfo
  {author} {\bibfnamefont {M.}~\bibnamefont {Kostylev}},\ and\ \bibinfo
  {author} {\bibfnamefont {M.}~\bibnamefont {Bailleul}},\ }\bibfield  {title}
  {\enquote {\bibinfo {title} {Frequency nonreciprocity of surface spin wave in
  permalloy thin films},}\ }\href {https://doi.org/10.1103/physrevb.93.054430}
  {\bibfield  {journal} {\bibinfo  {journal} {Phys. Rev. B}\ }\textbf {\bibinfo
  {volume} {93}},\ \bibinfo {pages} {54430} (\bibinfo {year}
  {2016})}\BibitemShut {NoStop}%
\bibitem [{\citenamefont {Sluka}\ \emph {et~al.}(2019)\citenamefont {Sluka},
  \citenamefont {Schneider}, \citenamefont {Gallardo}, \citenamefont
  {K{\'a}kay}, \citenamefont {Weigand}, \citenamefont {Warnatz}, \citenamefont
  {Mattheis}, \citenamefont {{Rold{\'a}n-Molina}}, \citenamefont {Landeros},
  \citenamefont {Tiberkevich}, \citenamefont {Slavin}, \citenamefont
  {Sch{\"u}tz}, \citenamefont {Erbe}, \citenamefont {Deac}, \citenamefont
  {Lindner}, \citenamefont {Raabe}, \citenamefont {Fassbender},\ and\
  \citenamefont {Wintz}}]{slukaEmissionPropagation1D2019a}%
  \BibitemOpen
  \bibfield  {author} {\bibinfo {author} {\bibfnamefont {V.}~\bibnamefont
  {Sluka}}, \bibinfo {author} {\bibfnamefont {T.}~\bibnamefont {Schneider}},
  \bibinfo {author} {\bibfnamefont {R.~A.}\ \bibnamefont {Gallardo}}, \bibinfo
  {author} {\bibfnamefont {A.}~\bibnamefont {K{\'a}kay}}, \bibinfo {author}
  {\bibfnamefont {M.}~\bibnamefont {Weigand}}, \bibinfo {author} {\bibfnamefont
  {T.}~\bibnamefont {Warnatz}}, \bibinfo {author} {\bibfnamefont
  {R.}~\bibnamefont {Mattheis}}, \bibinfo {author} {\bibfnamefont
  {A.}~\bibnamefont {{Rold{\'a}n-Molina}}}, \bibinfo {author} {\bibfnamefont
  {P.}~\bibnamefont {Landeros}}, \bibinfo {author} {\bibfnamefont
  {V.}~\bibnamefont {Tiberkevich}}, \bibinfo {author} {\bibfnamefont
  {A.}~\bibnamefont {Slavin}}, \bibinfo {author} {\bibfnamefont
  {G.}~\bibnamefont {Sch{\"u}tz}}, \bibinfo {author} {\bibfnamefont
  {A.}~\bibnamefont {Erbe}}, \bibinfo {author} {\bibfnamefont {A.}~\bibnamefont
  {Deac}}, \bibinfo {author} {\bibfnamefont {J.}~\bibnamefont {Lindner}},
  \bibinfo {author} {\bibfnamefont {J.}~\bibnamefont {Raabe}}, \bibinfo
  {author} {\bibfnamefont {J.}~\bibnamefont {Fassbender}},\ and\ \bibinfo
  {author} {\bibfnamefont {S.}~\bibnamefont {Wintz}},\ }\bibfield  {title}
  {\enquote {\bibinfo {title} {Emission and propagation of {{1D}} and {{2D}}
  spin waves with nanoscale wavelengths in anisotropic spin textures},}\ }\href
  {https://doi.org/10.1038/s41565-019-0383-4} {\bibfield  {journal} {\bibinfo
  {journal} {Nature Nanotechnology}\ }\textbf {\bibinfo {volume} {14}},\
  \bibinfo {pages} {328--333} (\bibinfo {year} {2019})}\BibitemShut {NoStop}%
\bibitem [{\citenamefont {Gallardo}\ \emph
  {et~al.}(2019{\natexlab{a}})\citenamefont {Gallardo}, \citenamefont
  {Schneider}, \citenamefont {Chaurasiya}, \citenamefont {Oelschl{\"a}gel},
  \citenamefont {Arekapudi}, \citenamefont {{Rold{\'a}n-Molina}}, \citenamefont
  {H{\"u}bner}, \citenamefont {Lenz}, \citenamefont {Barman}, \citenamefont
  {Fassbender}, \citenamefont {Lindner}, \citenamefont {Hellwig},\ and\
  \citenamefont
  {Landeros}}]{gallardoReconfigurableSpinWaveNonreciprocity2019a}%
  \BibitemOpen
  \bibfield  {author} {\bibinfo {author} {\bibfnamefont {R.}~\bibnamefont
  {Gallardo}}, \bibinfo {author} {\bibfnamefont {T.}~\bibnamefont {Schneider}},
  \bibinfo {author} {\bibfnamefont {A.}~\bibnamefont {Chaurasiya}}, \bibinfo
  {author} {\bibfnamefont {A.}~\bibnamefont {Oelschl{\"a}gel}}, \bibinfo
  {author} {\bibfnamefont {S.}~\bibnamefont {Arekapudi}}, \bibinfo {author}
  {\bibfnamefont {A.}~\bibnamefont {{Rold{\'a}n-Molina}}}, \bibinfo {author}
  {\bibfnamefont {R.}~\bibnamefont {H{\"u}bner}}, \bibinfo {author}
  {\bibfnamefont {K.}~\bibnamefont {Lenz}}, \bibinfo {author} {\bibfnamefont
  {A.}~\bibnamefont {Barman}}, \bibinfo {author} {\bibfnamefont
  {J.}~\bibnamefont {Fassbender}}, \bibinfo {author} {\bibfnamefont
  {J.}~\bibnamefont {Lindner}}, \bibinfo {author} {\bibfnamefont
  {O.}~\bibnamefont {Hellwig}},\ and\ \bibinfo {author} {\bibfnamefont
  {P.}~\bibnamefont {Landeros}},\ }\bibfield  {title} {\enquote {\bibinfo
  {title} {Reconfigurable {{Spin-Wave Nonreciprocity Induced}} by {{Dipolar
  Interaction}} in a {{Coupled Ferromagnetic Bilayer}}},}\ }\href
  {https://doi.org/10.1103/PhysRevApplied.12.034012} {\bibfield  {journal}
  {\bibinfo  {journal} {Physical Review Applied}\ }\textbf {\bibinfo {volume}
  {12}},\ \bibinfo {pages} {034012} (\bibinfo {year}
  {2019}{\natexlab{a}})}\BibitemShut {NoStop}%
\bibitem [{\citenamefont {Ishibashi}\ \emph {et~al.}(2020)\citenamefont
  {Ishibashi}, \citenamefont {Shiota}, \citenamefont {Li}, \citenamefont
  {Funada}, \citenamefont {Moriyama},\ and\ \citenamefont
  {Ono}}]{ishibashiSwitchableGiantNonreciprocal2020}%
  \BibitemOpen
  \bibfield  {author} {\bibinfo {author} {\bibfnamefont {M.}~\bibnamefont
  {Ishibashi}}, \bibinfo {author} {\bibfnamefont {Y.}~\bibnamefont {Shiota}},
  \bibinfo {author} {\bibfnamefont {T.}~\bibnamefont {Li}}, \bibinfo {author}
  {\bibfnamefont {S.}~\bibnamefont {Funada}}, \bibinfo {author} {\bibfnamefont
  {T.}~\bibnamefont {Moriyama}},\ and\ \bibinfo {author} {\bibfnamefont
  {T.}~\bibnamefont {Ono}},\ }\bibfield  {title} {\enquote {\bibinfo {title}
  {Switchable giant nonreciprocal frequency shift of propagating spin waves in
  synthetic antiferromagnets},}\ }\href
  {https://doi.org/10.1126/sciadv.aaz6931} {\bibfield  {journal} {\bibinfo
  {journal} {Science Advances}\ }\textbf {\bibinfo {volume} {6}},\ \bibinfo
  {pages} {eaaz6931} (\bibinfo {year} {2020})}\BibitemShut {NoStop}%
\bibitem [{\citenamefont {Albisetti}\ \emph {et~al.}(2020)\citenamefont
  {Albisetti}, \citenamefont {Tacchi}, \citenamefont {Silvani}, \citenamefont
  {Scaramuzzi}, \citenamefont {Finizio}, \citenamefont {Wintz}, \citenamefont
  {Rinaldi}, \citenamefont {Cantoni}, \citenamefont {Raabe}, \citenamefont
  {Carlotti}, \citenamefont {Bertacco}, \citenamefont {Riedo},\ and\
  \citenamefont {Petti}}]{albisettiOpticallyInspiredNanomagnonics2020}%
  \BibitemOpen
  \bibfield  {author} {\bibinfo {author} {\bibfnamefont {E.}~\bibnamefont
  {Albisetti}}, \bibinfo {author} {\bibfnamefont {S.}~\bibnamefont {Tacchi}},
  \bibinfo {author} {\bibfnamefont {R.}~\bibnamefont {Silvani}}, \bibinfo
  {author} {\bibfnamefont {G.}~\bibnamefont {Scaramuzzi}}, \bibinfo {author}
  {\bibfnamefont {S.}~\bibnamefont {Finizio}}, \bibinfo {author} {\bibfnamefont
  {S.}~\bibnamefont {Wintz}}, \bibinfo {author} {\bibfnamefont
  {C.}~\bibnamefont {Rinaldi}}, \bibinfo {author} {\bibfnamefont
  {M.}~\bibnamefont {Cantoni}}, \bibinfo {author} {\bibfnamefont
  {J.}~\bibnamefont {Raabe}}, \bibinfo {author} {\bibfnamefont
  {G.}~\bibnamefont {Carlotti}}, \bibinfo {author} {\bibfnamefont
  {R.}~\bibnamefont {Bertacco}}, \bibinfo {author} {\bibfnamefont
  {E.}~\bibnamefont {Riedo}},\ and\ \bibinfo {author} {\bibfnamefont
  {D.}~\bibnamefont {Petti}},\ }\bibfield  {title} {\enquote {\bibinfo {title}
  {Optically {{Inspired Nanomagnonics}} with {{Nonreciprocal Spin Waves}} in
  {{Synthetic Antiferromagnets}}},}\ }\href
  {https://doi.org/10.1002/adma.201906439} {\bibfield  {journal} {\bibinfo
  {journal} {Advanced Materials}\ }\textbf {\bibinfo {volume} {32}},\ \bibinfo
  {pages} {1906439} (\bibinfo {year} {2020})}\BibitemShut {NoStop}%
\bibitem [{\citenamefont {Grassi}\ \emph {et~al.}(2020)\citenamefont {Grassi},
  \citenamefont {Geilen}, \citenamefont {Louis}, \citenamefont {Mohseni},
  \citenamefont {Br{\"a}cher}, \citenamefont {Hehn}, \citenamefont {Stoeffler},
  \citenamefont {Bailleul}, \citenamefont {Pirro},\ and\ \citenamefont
  {Henry}}]{grassiSlowWaveBasedNanomagnonicDiode2020}%
  \BibitemOpen
  \bibfield  {author} {\bibinfo {author} {\bibfnamefont {M.}~\bibnamefont
  {Grassi}}, \bibinfo {author} {\bibfnamefont {M.}~\bibnamefont {Geilen}},
  \bibinfo {author} {\bibfnamefont {D.}~\bibnamefont {Louis}}, \bibinfo
  {author} {\bibfnamefont {M.}~\bibnamefont {Mohseni}}, \bibinfo {author}
  {\bibfnamefont {T.}~\bibnamefont {Br{\"a}cher}}, \bibinfo {author}
  {\bibfnamefont {M.}~\bibnamefont {Hehn}}, \bibinfo {author} {\bibfnamefont
  {D.}~\bibnamefont {Stoeffler}}, \bibinfo {author} {\bibfnamefont
  {M.}~\bibnamefont {Bailleul}}, \bibinfo {author} {\bibfnamefont
  {P.}~\bibnamefont {Pirro}},\ and\ \bibinfo {author} {\bibfnamefont
  {Y.}~\bibnamefont {Henry}},\ }\bibfield  {title} {\enquote {\bibinfo {title}
  {Slow-{{Wave-Based Nanomagnonic Diode}}},}\ }\href
  {https://doi.org/10.1103/PhysRevApplied.14.024047} {\bibfield  {journal}
  {\bibinfo  {journal} {Physical Review Applied}\ }\textbf {\bibinfo {volume}
  {14}},\ \bibinfo {pages} {024047} (\bibinfo {year} {2020})}\BibitemShut
  {NoStop}%
\bibitem [{\citenamefont {Gallardo}\ \emph {et~al.}(2021)\citenamefont
  {Gallardo}, \citenamefont {{Alvarado-Seguel}}, \citenamefont {K{\'a}kay},
  \citenamefont {Lindner},\ and\ \citenamefont
  {Landeros}}]{gallardoSpinwaveFocusingInduced2021}%
  \BibitemOpen
  \bibfield  {author} {\bibinfo {author} {\bibfnamefont {R.~A.}\ \bibnamefont
  {Gallardo}}, \bibinfo {author} {\bibfnamefont {P.}~\bibnamefont
  {{Alvarado-Seguel}}}, \bibinfo {author} {\bibfnamefont {A.}~\bibnamefont
  {K{\'a}kay}}, \bibinfo {author} {\bibfnamefont {J.}~\bibnamefont {Lindner}},\
  and\ \bibinfo {author} {\bibfnamefont {P.}~\bibnamefont {Landeros}},\
  }\bibfield  {title} {\enquote {\bibinfo {title} {Spin-wave focusing induced
  by dipole-dipole interaction in synthetic antiferromagnets},}\ }\href
  {https://doi.org/10.1103/PhysRevB.104.174417} {\bibfield  {journal} {\bibinfo
   {journal} {Physical Review B}\ }\textbf {\bibinfo {volume} {104}},\ \bibinfo
  {pages} {174417} (\bibinfo {year} {2021})}\BibitemShut {NoStop}%
\bibitem [{\citenamefont {Zhang}\ and\ \citenamefont
  {Tchernyshyov}(2018)}]{zhangFerromagneticDomainWall2018}%
  \BibitemOpen
  \bibfield  {author} {\bibinfo {author} {\bibfnamefont {S.}~\bibnamefont
  {Zhang}}\ and\ \bibinfo {author} {\bibfnamefont {O.}~\bibnamefont
  {Tchernyshyov}},\ }\bibfield  {title} {\enquote {\bibinfo {title}
  {Ferromagnetic domain wall as a nonreciprocal string},}\ }\href@noop {}
  {\bibfield  {journal} {\bibinfo  {journal} {arXiv e-prints}\ } (\bibinfo
  {year} {2018})}\BibitemShut {NoStop}%
\bibitem [{\citenamefont {Henry}\ \emph {et~al.}(2019)\citenamefont {Henry},
  \citenamefont {Stoeffler}, \citenamefont {Kim},\ and\ \citenamefont
  {Bailleul}}]{henryUnidirectionalSpinwaveChanneling2019}%
  \BibitemOpen
  \bibfield  {author} {\bibinfo {author} {\bibfnamefont {Y.}~\bibnamefont
  {Henry}}, \bibinfo {author} {\bibfnamefont {D.}~\bibnamefont {Stoeffler}},
  \bibinfo {author} {\bibfnamefont {J.-V.}\ \bibnamefont {Kim}},\ and\ \bibinfo
  {author} {\bibfnamefont {M.}~\bibnamefont {Bailleul}},\ }\bibfield  {title}
  {\enquote {\bibinfo {title} {Unidirectional spin-wave channeling along
  magnetic domain walls of {{Bloch}} type},}\ }\href
  {https://doi.org/10.1103/PhysRevB.100.024416} {\bibfield  {journal} {\bibinfo
   {journal} {Physical Review B}\ }\textbf {\bibinfo {volume} {100}},\ \bibinfo
  {pages} {024416} (\bibinfo {year} {2019})}\BibitemShut {NoStop}%
\bibitem [{\citenamefont {K{\"o}rber}\ \emph {et~al.}(2017)\citenamefont
  {K{\"o}rber}, \citenamefont {Wagner}, \citenamefont {K{\'a}kay},\ and\
  \citenamefont {Schultheiss}}]{korberSpinWaveReciprocityPresence2017}%
  \BibitemOpen
  \bibfield  {author} {\bibinfo {author} {\bibfnamefont {L.}~\bibnamefont
  {K{\"o}rber}}, \bibinfo {author} {\bibfnamefont {K.}~\bibnamefont {Wagner}},
  \bibinfo {author} {\bibfnamefont {A.}~\bibnamefont {K{\'a}kay}},\ and\
  \bibinfo {author} {\bibfnamefont {H.}~\bibnamefont {Schultheiss}},\
  }\bibfield  {title} {\enquote {\bibinfo {title} {Spin-{{Wave Reciprocity}} in
  the {{Presence}} of {{N\'eel Walls}}},}\ }\href
  {https://doi.org/10.1109/LMAG.2017.2762642} {\bibfield  {journal} {\bibinfo
  {journal} {IEEE Magnetics Letters}\ }\textbf {\bibinfo {volume} {8}},\
  \bibinfo {pages} {1--4} (\bibinfo {year} {2017})}\BibitemShut {NoStop}%
\bibitem [{\citenamefont {Spaldin}, \citenamefont {Fiebig},\ and\ \citenamefont
  {Mostovoy}(2008)}]{spaldinToroidalMomentCondensedmatter2008a}%
  \BibitemOpen
  \bibfield  {author} {\bibinfo {author} {\bibfnamefont {N.~A.}\ \bibnamefont
  {Spaldin}}, \bibinfo {author} {\bibfnamefont {M.}~\bibnamefont {Fiebig}},\
  and\ \bibinfo {author} {\bibfnamefont {M.}~\bibnamefont {Mostovoy}},\
  }\bibfield  {title} {\enquote {\bibinfo {title} {The toroidal moment in
  condensed-matter physics and its relation to the magnetoelectric effect},}\
  }\href {https://doi.org/10.1088/0953-8984/20/43/434203} {\bibfield  {journal}
  {\bibinfo  {journal} {Journal of Physics: Condensed Matter}\ }\textbf
  {\bibinfo {volume} {20}},\ \bibinfo {pages} {434203} (\bibinfo {year}
  {2008})}\BibitemShut {NoStop}%
\bibitem [{\citenamefont {K{\'e}zsm{\'a}rki}\ \emph
  {et~al.}(2014{\natexlab{b}})\citenamefont {K{\'e}zsm{\'a}rki}, \citenamefont
  {Szaller}, \citenamefont {Bord{\'a}cs}, \citenamefont {Kocsis}, \citenamefont
  {Tokunaga}, \citenamefont {Taguchi}, \citenamefont {Murakawa}, \citenamefont
  {Tokura}, \citenamefont {Engelkamp}, \citenamefont {R{\~o}{\~o}m},\ and\
  \citenamefont {Nagel}}]{kezsmarkiOnewayTransparencyFourcoloured2014d}%
  \BibitemOpen
  \bibfield  {author} {\bibinfo {author} {\bibfnamefont {I.}~\bibnamefont
  {K{\'e}zsm{\'a}rki}}, \bibinfo {author} {\bibfnamefont {D.}~\bibnamefont
  {Szaller}}, \bibinfo {author} {\bibfnamefont {S.}~\bibnamefont
  {Bord{\'a}cs}}, \bibinfo {author} {\bibfnamefont {V.}~\bibnamefont {Kocsis}},
  \bibinfo {author} {\bibfnamefont {Y.}~\bibnamefont {Tokunaga}}, \bibinfo
  {author} {\bibfnamefont {Y.}~\bibnamefont {Taguchi}}, \bibinfo {author}
  {\bibfnamefont {H.}~\bibnamefont {Murakawa}}, \bibinfo {author}
  {\bibfnamefont {Y.}~\bibnamefont {Tokura}}, \bibinfo {author} {\bibfnamefont
  {H.}~\bibnamefont {Engelkamp}}, \bibinfo {author} {\bibfnamefont
  {T.}~\bibnamefont {R{\~o}{\~o}m}},\ and\ \bibinfo {author} {\bibfnamefont
  {U.}~\bibnamefont {Nagel}},\ }\bibfield  {title} {\enquote {\bibinfo {title}
  {One-way transparency of four-coloured spin-wave excitations in multiferroic
  materials},}\ }\href {https://doi.org/10.1038/ncomms4203} {\bibfield
  {journal} {\bibinfo  {journal} {Nature Communications}\ }\textbf {\bibinfo
  {volume} {5}},\ \bibinfo {pages} {3203} (\bibinfo {year}
  {2014}{\natexlab{b}})}\BibitemShut {NoStop}%
\bibitem [{\citenamefont {K{\'e}zsm{\'a}rki}\ \emph {et~al.}(2011)\citenamefont
  {K{\'e}zsm{\'a}rki}, \citenamefont {Kida}, \citenamefont {Murakawa},
  \citenamefont {Bord{\'a}cs}, \citenamefont {Onose},\ and\ \citenamefont
  {Tokura}}]{kezsmarkiEnhancedDirectionalDichroism2011b}%
  \BibitemOpen
  \bibfield  {author} {\bibinfo {author} {\bibfnamefont {I.}~\bibnamefont
  {K{\'e}zsm{\'a}rki}}, \bibinfo {author} {\bibfnamefont {N.}~\bibnamefont
  {Kida}}, \bibinfo {author} {\bibfnamefont {H.}~\bibnamefont {Murakawa}},
  \bibinfo {author} {\bibfnamefont {S.}~\bibnamefont {Bord{\'a}cs}}, \bibinfo
  {author} {\bibfnamefont {Y.}~\bibnamefont {Onose}},\ and\ \bibinfo {author}
  {\bibfnamefont {Y.}~\bibnamefont {Tokura}},\ }\bibfield  {title} {\enquote
  {\bibinfo {title} {Enhanced {{Directional Dichroism}} of {{Terahertz Light}}
  in {{Resonance}} with {{Magnetic Excitations}} of the {{Multiferroic Ba}} 2
  {{CoGe}} 2 {{O}} 7 {{Oxide Compound}}},}\ }\href
  {https://doi.org/10.1103/PhysRevLett.106.057403} {\bibfield  {journal}
  {\bibinfo  {journal} {Physical Review Letters}\ }\textbf {\bibinfo {volume}
  {106}},\ \bibinfo {pages} {057403} (\bibinfo {year} {2011})}\BibitemShut
  {NoStop}%
\bibitem [{\citenamefont {Szaller}, \citenamefont {Bord{\'a}cs},\ and\
  \citenamefont
  {K{\'e}zsm{\'a}rki}(2013)}]{szallerSymmetryConditionsNonreciprocal2013b}%
  \BibitemOpen
  \bibfield  {author} {\bibinfo {author} {\bibfnamefont {D.}~\bibnamefont
  {Szaller}}, \bibinfo {author} {\bibfnamefont {S.}~\bibnamefont
  {Bord{\'a}cs}},\ and\ \bibinfo {author} {\bibfnamefont {I.}~\bibnamefont
  {K{\'e}zsm{\'a}rki}},\ }\bibfield  {title} {\enquote {\bibinfo {title}
  {Symmetry conditions for nonreciprocal light propagation in magnetic
  crystals},}\ }\href {https://doi.org/10.1103/PhysRevB.87.014421} {\bibfield
  {journal} {\bibinfo  {journal} {Physical Review B}\ }\textbf {\bibinfo
  {volume} {87}},\ \bibinfo {pages} {014421} (\bibinfo {year}
  {2013})}\BibitemShut {NoStop}%
\bibitem [{\citenamefont {Okamura}\ \emph {et~al.}(2013)\citenamefont
  {Okamura}, \citenamefont {Kagawa}, \citenamefont {Mochizuki}, \citenamefont
  {Kubota}, \citenamefont {Seki}, \citenamefont {Ishiwata}, \citenamefont
  {Kawasaki}, \citenamefont {Onose},\ and\ \citenamefont
  {Tokura}}]{okamuraMicrowaveMagnetoelectricEffect2013}%
  \BibitemOpen
  \bibfield  {author} {\bibinfo {author} {\bibfnamefont {Y.}~\bibnamefont
  {Okamura}}, \bibinfo {author} {\bibfnamefont {F.}~\bibnamefont {Kagawa}},
  \bibinfo {author} {\bibfnamefont {M.}~\bibnamefont {Mochizuki}}, \bibinfo
  {author} {\bibfnamefont {M.}~\bibnamefont {Kubota}}, \bibinfo {author}
  {\bibfnamefont {S.}~\bibnamefont {Seki}}, \bibinfo {author} {\bibfnamefont
  {S.}~\bibnamefont {Ishiwata}}, \bibinfo {author} {\bibfnamefont
  {M.}~\bibnamefont {Kawasaki}}, \bibinfo {author} {\bibfnamefont
  {Y.}~\bibnamefont {Onose}},\ and\ \bibinfo {author} {\bibfnamefont
  {Y.}~\bibnamefont {Tokura}},\ }\bibfield  {title} {\enquote {\bibinfo {title}
  {Microwave magnetoelectric effect via skyrmion resonance modes in a
  helimagnetic multiferroic},}\ }\href {https://doi.org/10.1038/ncomms3391}
  {\bibfield  {journal} {\bibinfo  {journal} {Nature Communications}\ }\textbf
  {\bibinfo {volume} {4}},\ \bibinfo {pages} {2391} (\bibinfo {year}
  {2013})}\BibitemShut {NoStop}%
\bibitem [{\citenamefont {Kocsis}\ \emph {et~al.}(2018)\citenamefont {Kocsis},
  \citenamefont {Penc}, \citenamefont {R{\~o}{\~o}m}, \citenamefont {Nagel},
  \citenamefont {V{\'i}t}, \citenamefont {Romh{\'a}nyi}, \citenamefont
  {Tokunaga}, \citenamefont {Taguchi}, \citenamefont {Tokura}, \citenamefont
  {K{\'e}zsm{\'a}rki},\ and\ \citenamefont
  {Bord{\'a}cs}}]{kocsisIdentificationAntiferromagneticDomains2018}%
  \BibitemOpen
  \bibfield  {author} {\bibinfo {author} {\bibfnamefont {V.}~\bibnamefont
  {Kocsis}}, \bibinfo {author} {\bibfnamefont {K.}~\bibnamefont {Penc}},
  \bibinfo {author} {\bibfnamefont {T.}~\bibnamefont {R{\~o}{\~o}m}}, \bibinfo
  {author} {\bibfnamefont {U.}~\bibnamefont {Nagel}}, \bibinfo {author}
  {\bibfnamefont {J.}~\bibnamefont {V{\'i}t}}, \bibinfo {author} {\bibfnamefont
  {J.}~\bibnamefont {Romh{\'a}nyi}}, \bibinfo {author} {\bibfnamefont
  {Y.}~\bibnamefont {Tokunaga}}, \bibinfo {author} {\bibfnamefont
  {Y.}~\bibnamefont {Taguchi}}, \bibinfo {author} {\bibfnamefont
  {Y.}~\bibnamefont {Tokura}}, \bibinfo {author} {\bibfnamefont
  {I.}~\bibnamefont {K{\'e}zsm{\'a}rki}},\ and\ \bibinfo {author}
  {\bibfnamefont {S.}~\bibnamefont {Bord{\'a}cs}},\ }\bibfield  {title}
  {\enquote {\bibinfo {title} {Identification of {{Antiferromagnetic Domains
  Via}} the {{Optical Magnetoelectric Effect}}},}\ }\href
  {https://doi.org/10.1103/PhysRevLett.121.057601} {\bibfield  {journal}
  {\bibinfo  {journal} {Physical Review Letters}\ }\textbf {\bibinfo {volume}
  {121}},\ \bibinfo {pages} {057601} (\bibinfo {year} {2018})}\BibitemShut
  {NoStop}%
\bibitem [{\citenamefont {K{\"o}rber}\ \emph {et~al.}(2021)\citenamefont
  {K{\"o}rber}, \citenamefont {Quasebarth}, \citenamefont {Otto},\ and\
  \citenamefont {K{\'a}kay}}]{korberFiniteelementDynamicmatrixApproach2021}%
  \BibitemOpen
  \bibfield  {author} {\bibinfo {author} {\bibfnamefont {L.}~\bibnamefont
  {K{\"o}rber}}, \bibinfo {author} {\bibfnamefont {G.}~\bibnamefont
  {Quasebarth}}, \bibinfo {author} {\bibfnamefont {A.}~\bibnamefont {Otto}},\
  and\ \bibinfo {author} {\bibfnamefont {A.}~\bibnamefont {K{\'a}kay}},\
  }\bibfield  {title} {\enquote {\bibinfo {title} {Finite-element
  dynamic-matrix approach for spin-wave dispersions in magnonic waveguides with
  arbitrary cross section},}\ }\href {https://doi.org/10.1063/5.0054169}
  {\bibfield  {journal} {\bibinfo  {journal} {AIP Advances}\ }\textbf {\bibinfo
  {volume} {11}},\ \bibinfo {pages} {095006} (\bibinfo {year}
  {2021})}\BibitemShut {NoStop}%
\bibitem [{\citenamefont {Ot{\'a}lora}\ \emph {et~al.}(2018)\citenamefont
  {Ot{\'a}lora}, \citenamefont {K{\'a}kay}, \citenamefont {Lindner},
  \citenamefont {Schultheiss}, \citenamefont {Thomas}, \citenamefont
  {Fassbender},\ and\ \citenamefont
  {Nielsch}}]{otaloraFrequencyLinewidthDecay2018}%
  \BibitemOpen
  \bibfield  {author} {\bibinfo {author} {\bibfnamefont {J.~A.}\ \bibnamefont
  {Ot{\'a}lora}}, \bibinfo {author} {\bibfnamefont {A.}~\bibnamefont
  {K{\'a}kay}}, \bibinfo {author} {\bibfnamefont {J.}~\bibnamefont {Lindner}},
  \bibinfo {author} {\bibfnamefont {H.}~\bibnamefont {Schultheiss}}, \bibinfo
  {author} {\bibfnamefont {A.}~\bibnamefont {Thomas}}, \bibinfo {author}
  {\bibfnamefont {J.}~\bibnamefont {Fassbender}},\ and\ \bibinfo {author}
  {\bibfnamefont {K.}~\bibnamefont {Nielsch}},\ }\bibfield  {title} {\enquote
  {\bibinfo {title} {Frequency linewidth and decay length of spin waves in
  curved magnetic membranes},}\ }\href
  {https://doi.org/10.1103/PhysRevB.98.014403} {\bibfield  {journal} {\bibinfo
  {journal} {Physical Review B}\ }\textbf {\bibinfo {volume} {98}},\ \bibinfo
  {pages} {014403} (\bibinfo {year} {2018})}\BibitemShut {NoStop}%
\bibitem [{\citenamefont {{Salazar-Cardona}}\ \emph {et~al.}(2021)\citenamefont
  {{Salazar-Cardona}}, \citenamefont {K{\"o}rber}, \citenamefont {Schultheiss},
  \citenamefont {Lenz}, \citenamefont {Thomas}, \citenamefont {Nielsch},
  \citenamefont {K{\'a}kay},\ and\ \citenamefont
  {Ot{\'a}lora}}]{salazar-cardonaNonreciprocitySpinWaves2021}%
  \BibitemOpen
  \bibfield  {author} {\bibinfo {author} {\bibfnamefont {M.~M.}\ \bibnamefont
  {{Salazar-Cardona}}}, \bibinfo {author} {\bibfnamefont {L.}~\bibnamefont
  {K{\"o}rber}}, \bibinfo {author} {\bibfnamefont {H.}~\bibnamefont
  {Schultheiss}}, \bibinfo {author} {\bibfnamefont {K.}~\bibnamefont {Lenz}},
  \bibinfo {author} {\bibfnamefont {A.}~\bibnamefont {Thomas}}, \bibinfo
  {author} {\bibfnamefont {K.}~\bibnamefont {Nielsch}}, \bibinfo {author}
  {\bibfnamefont {A.}~\bibnamefont {K{\'a}kay}},\ and\ \bibinfo {author}
  {\bibfnamefont {J.~A.}\ \bibnamefont {Ot{\'a}lora}},\ }\bibfield  {title}
  {\enquote {\bibinfo {title} {Nonreciprocity of spin waves in magnetic
  nanotubes with helical equilibrium magnetization},}\ }\href
  {https://doi.org/10.1063/5.0048692} {\bibfield  {journal} {\bibinfo
  {journal} {Applied Physics Letters}\ }\textbf {\bibinfo {volume} {118}},\
  \bibinfo {pages} {262411} (\bibinfo {year} {2021})}\BibitemShut {NoStop}%
\bibitem [{\citenamefont {Landeros}\ \emph {et~al.}(2009)\citenamefont
  {Landeros}, \citenamefont {Suarez}, \citenamefont {Cuchillo},\ and\
  \citenamefont {Vargas}}]{landerosEquilibriumStatesVortex2009}%
  \BibitemOpen
  \bibfield  {author} {\bibinfo {author} {\bibfnamefont {P.}~\bibnamefont
  {Landeros}}, \bibinfo {author} {\bibfnamefont {O.~J.}\ \bibnamefont
  {Suarez}}, \bibinfo {author} {\bibfnamefont {A.}~\bibnamefont {Cuchillo}},\
  and\ \bibinfo {author} {\bibfnamefont {P.}~\bibnamefont {Vargas}},\
  }\bibfield  {title} {\enquote {\bibinfo {title} {Equilibrium states and
  vortex domain wall nucleation in ferromagnetic nanotubes},}\ }\href
  {https://doi.org/10.1103/PhysRevB.79.024404} {\bibfield  {journal} {\bibinfo
  {journal} {Phys. Rev. B}\ }\textbf {\bibinfo {volume} {79}},\ \bibinfo
  {pages} {24404} (\bibinfo {year} {2009})}\BibitemShut {NoStop}%
\bibitem [{\citenamefont {Ot{\'a}lora}\ \emph {et~al.}(2015)\citenamefont
  {Ot{\'a}lora}, \citenamefont {{Cort{\'e}s-Ortu{\~n}o}}, \citenamefont
  {G{\"o}rlitz}, \citenamefont {Nielsch},\ and\ \citenamefont
  {Landeros}}]{otaloraOerstedFieldAssisted2015}%
  \BibitemOpen
  \bibfield  {author} {\bibinfo {author} {\bibfnamefont {J.~A.}\ \bibnamefont
  {Ot{\'a}lora}}, \bibinfo {author} {\bibfnamefont {D.}~\bibnamefont
  {{Cort{\'e}s-Ortu{\~n}o}}}, \bibinfo {author} {\bibfnamefont
  {D.}~\bibnamefont {G{\"o}rlitz}}, \bibinfo {author} {\bibfnamefont
  {K.}~\bibnamefont {Nielsch}},\ and\ \bibinfo {author} {\bibfnamefont
  {P.}~\bibnamefont {Landeros}},\ }\bibfield  {title} {\enquote {\bibinfo
  {title} {Oersted field assisted magnetization reversal in cylindrical
  core-shell nanostructures},}\ }\href {https://doi.org/10.1063/1.4919746}
  {\bibfield  {journal} {\bibinfo  {journal} {Journal of Applied Physics}\
  }\textbf {\bibinfo {volume} {117}},\ \bibinfo {pages} {173914} (\bibinfo
  {year} {2015})}\BibitemShut {NoStop}%
\bibitem [{Note1()}]{Note1}%
  \BibitemOpen
  \bibinfo {note} {The critical field in Eq.~\ref {eq:critical-field} can be
  obtained from Ref.~\protect \citenum
  {salazar-cardonaNonreciprocitySpinWaves2021} by replacing inner and outer
  radii $r_{1,2}$ with thickness and average radius as $r_{1,2}=R\mp
  T/2$.}\BibitemShut {Stop}%
\bibitem [{\citenamefont {K{\"o}rber}\ \emph {et~al.}(2022)\citenamefont
  {K{\"o}rber}, \citenamefont {Quasebarth}, \citenamefont {Hempel},
  \citenamefont {Zahn}, \citenamefont {Andreas}, \citenamefont {Westphal},
  \citenamefont {Hertel},\ and\ \citenamefont
  {K{\'a}kay}}]{korberTetraXFiniteElementMicromagneticModeling2022}%
  \BibitemOpen
  \bibfield  {author} {\bibinfo {author} {\bibfnamefont {L.}~\bibnamefont
  {K{\"o}rber}}, \bibinfo {author} {\bibfnamefont {G.}~\bibnamefont
  {Quasebarth}}, \bibinfo {author} {\bibfnamefont {A.}~\bibnamefont {Hempel}},
  \bibinfo {author} {\bibfnamefont {F.}~\bibnamefont {Zahn}}, \bibinfo {author}
  {\bibfnamefont {O.}~\bibnamefont {Andreas}}, \bibinfo {author} {\bibfnamefont
  {E.}~\bibnamefont {Westphal}}, \bibinfo {author} {\bibfnamefont
  {R.}~\bibnamefont {Hertel}},\ and\ \bibinfo {author} {\bibfnamefont
  {A.}~\bibnamefont {K{\'a}kay}},\ }\href
  {https://doi.org/10.14278/rodare.1418} {\enquote {\bibinfo {title}
  {{{TetraX}}: {{Finite-Element Micromagnetic-Modeling Package}}},}\ }
  (\bibinfo {year} {2022})\BibitemShut {NoStop}%
\bibitem [{\citenamefont {Fredkin}\ and\ \citenamefont
  {Koehler}(1990)}]{fredkinHybridMethodComputing1990}%
  \BibitemOpen
  \bibfield  {author} {\bibinfo {author} {\bibfnamefont {D.}~\bibnamefont
  {Fredkin}}\ and\ \bibinfo {author} {\bibfnamefont {T.}~\bibnamefont
  {Koehler}},\ }\bibfield  {title} {\enquote {\bibinfo {title} {Hybrid method
  for computing demagnetizing fields},}\ }\href
  {https://doi.org/10.1109/20.106342} {\bibfield  {journal} {\bibinfo
  {journal} {IEEE Transactions on Magnetics}\ }\textbf {\bibinfo {volume}
  {26}},\ \bibinfo {pages} {415--417} (\bibinfo {year} {1990})}\BibitemShut
  {NoStop}%
\bibitem [{\citenamefont {Lanczos}(1950)}]{lanczosIterationMethodSolution1950}%
  \BibitemOpen
  \bibfield  {author} {\bibinfo {author} {\bibfnamefont {C.}~\bibnamefont
  {Lanczos}},\ }\href@noop {} {\emph {\bibinfo {title} {An Iteration Method for
  the Solution of the Eigenvalue Problem of Linear Differential and Integral
  Operators}}}\ (\bibinfo  {publisher} {{United States Governm. Press Office
  Los Angeles, CA}},\ \bibinfo {year} {1950})\BibitemShut {NoStop}%
\bibitem [{\citenamefont {Arnoldi}(1951)}]{arnoldi1951principle}%
  \BibitemOpen
  \bibfield  {author} {\bibinfo {author} {\bibfnamefont {W.~E.}\ \bibnamefont
  {Arnoldi}},\ }\bibfield  {title} {\enquote {\bibinfo {title} {The principle
  of minimized iterations in the solution of the matrix eigenvalue problem},}\
  }\href@noop {} {\bibfield  {journal} {\bibinfo  {journal} {Quarterly of
  applied mathematics}\ }\textbf {\bibinfo {volume} {9}},\ \bibinfo {pages}
  {17--29} (\bibinfo {year} {1951})}\BibitemShut {NoStop}%
\bibitem [{\citenamefont {Ivanov}\ and\ \citenamefont
  {Zaspel}(2002)}]{ivanovMagnonModesThin2002}%
  \BibitemOpen
  \bibfield  {author} {\bibinfo {author} {\bibfnamefont {B.~A.}\ \bibnamefont
  {Ivanov}}\ and\ \bibinfo {author} {\bibfnamefont {C.~E.}\ \bibnamefont
  {Zaspel}},\ }\bibfield  {title} {\enquote {\bibinfo {title} {Magnon modes for
  thin circular vortex-state magnetic dots},}\ }\href
  {https://doi.org/10.1063/1.1499515} {\bibfield  {journal} {\bibinfo
  {journal} {Applied Physics Letters}\ }\textbf {\bibinfo {volume} {81}},\
  \bibinfo {pages} {1261--1263} (\bibinfo {year} {2002})}\BibitemShut {NoStop}%
\bibitem [{\citenamefont {Buess}\ \emph {et~al.}(2004)\citenamefont {Buess},
  \citenamefont {H{\textbackslash}uollinger}, \citenamefont {Haug},
  \citenamefont {Perzlmaier}, \citenamefont {Pescia}, \citenamefont
  {Scheinfein}, \citenamefont {{D.Weiss}},\ and\ \citenamefont
  {Back}}]{buessFourierTransformImaging2004a}%
  \BibitemOpen
  \bibfield  {author} {\bibinfo {author} {\bibfnamefont {M.}~\bibnamefont
  {Buess}}, \bibinfo {author} {\bibfnamefont {R.}~\bibnamefont
  {H{\textbackslash}uollinger}}, \bibinfo {author} {\bibfnamefont
  {T.}~\bibnamefont {Haug}}, \bibinfo {author} {\bibfnamefont {K.}~\bibnamefont
  {Perzlmaier}}, \bibinfo {author} {\bibfnamefont {U.~K.~D.}\ \bibnamefont
  {Pescia}}, \bibinfo {author} {\bibfnamefont {M.~R.}\ \bibnamefont
  {Scheinfein}}, \bibinfo {author} {\bibnamefont {{D.Weiss}}},\ and\ \bibinfo
  {author} {\bibfnamefont {C.~H.}\ \bibnamefont {Back}},\ }\bibfield  {title}
  {\enquote {\bibinfo {title} {Fourier transform imaging of spin vortex
  eigenmodes},}\ }\href {https://doi.org/10.1103/PhysRevLett.93.077207}
  {\bibfield  {journal} {\bibinfo  {journal} {Phys. Rev. Lett.}\ }\textbf
  {\bibinfo {volume} {93}},\ \bibinfo {pages} {77207} (\bibinfo {year}
  {2004})}\BibitemShut {NoStop}%
\bibitem [{\citenamefont {Zaspel}\ \emph {et~al.}(2005)\citenamefont {Zaspel},
  \citenamefont {Ivanov}, \citenamefont {Park},\ and\ \citenamefont
  {Crowell}}]{zaspelExcitationsVortexstatePermalloy2005}%
  \BibitemOpen
  \bibfield  {author} {\bibinfo {author} {\bibfnamefont {C.~E.}\ \bibnamefont
  {Zaspel}}, \bibinfo {author} {\bibfnamefont {B.~A.}\ \bibnamefont {Ivanov}},
  \bibinfo {author} {\bibfnamefont {J.~P.}\ \bibnamefont {Park}},\ and\
  \bibinfo {author} {\bibfnamefont {P.~A.}\ \bibnamefont {Crowell}},\
  }\bibfield  {title} {\enquote {\bibinfo {title} {Excitations in vortex-state
  permalloy dots},}\ }\href {https://doi.org/10.1103/PhysRevB.72.024427}
  {\bibfield  {journal} {\bibinfo  {journal} {Phys. Rev. B}\ }\textbf {\bibinfo
  {volume} {72}},\ \bibinfo {pages} {24427} (\bibinfo {year}
  {2005})}\BibitemShut {NoStop}%
\bibitem [{\citenamefont {Gaididei}\ \emph {et~al.}(2010)\citenamefont
  {Gaididei}, \citenamefont {Kravchuk}, \citenamefont {Sheka},\ and\
  \citenamefont {Mertens}}]{gaidideiMultipleVortexantivortexPair2010}%
  \BibitemOpen
  \bibfield  {author} {\bibinfo {author} {\bibfnamefont {Y.}~\bibnamefont
  {Gaididei}}, \bibinfo {author} {\bibfnamefont {V.~P.}\ \bibnamefont
  {Kravchuk}}, \bibinfo {author} {\bibfnamefont {D.~D.}\ \bibnamefont
  {Sheka}},\ and\ \bibinfo {author} {\bibfnamefont {F.~G.}\ \bibnamefont
  {Mertens}},\ }\bibfield  {title} {\enquote {\bibinfo {title} {Multiple
  vortex-antivortex pair generation in magnetic nanodots},}\ }\href
  {https://doi.org/10.1103/PhysRevB.81.094431} {\bibfield  {journal} {\bibinfo
  {journal} {Phys. Rev. B}\ }\textbf {\bibinfo {volume} {81}},\ \bibinfo
  {pages} {94431} (\bibinfo {year} {2010})}\BibitemShut {NoStop}%
\bibitem [{\citenamefont {Schultheiss}\ \emph {et~al.}()\citenamefont
  {Schultheiss}, \citenamefont {Verba}, \citenamefont {Wehrmann}, \citenamefont
  {Wagner}, \citenamefont {K{\"o}rber}, \citenamefont {Hula}, \citenamefont
  {Hache}, \citenamefont {Kakay}, \citenamefont {Awad}, \citenamefont
  {Tiberkevich}, \citenamefont {Slavin}, \citenamefont {Fassbender},\ and\
  \citenamefont {Schultheiss}}]{schultheissExcitationWhisperingGallery}%
  \BibitemOpen
  \bibfield  {author} {\bibinfo {author} {\bibfnamefont {K.}~\bibnamefont
  {Schultheiss}}, \bibinfo {author} {\bibfnamefont {R.}~\bibnamefont {Verba}},
  \bibinfo {author} {\bibfnamefont {F.}~\bibnamefont {Wehrmann}}, \bibinfo
  {author} {\bibfnamefont {K.}~\bibnamefont {Wagner}}, \bibinfo {author}
  {\bibfnamefont {L.}~\bibnamefont {K{\"o}rber}}, \bibinfo {author}
  {\bibfnamefont {T.}~\bibnamefont {Hula}}, \bibinfo {author} {\bibfnamefont
  {T.}~\bibnamefont {Hache}}, \bibinfo {author} {\bibfnamefont
  {A.}~\bibnamefont {Kakay}}, \bibinfo {author} {\bibfnamefont {A.~A.}\
  \bibnamefont {Awad}}, \bibinfo {author} {\bibfnamefont {V.}~\bibnamefont
  {Tiberkevich}}, \bibinfo {author} {\bibfnamefont {A.~N.}\ \bibnamefont
  {Slavin}}, \bibinfo {author} {\bibfnamefont {J.}~\bibnamefont {Fassbender}},\
  and\ \bibinfo {author} {\bibfnamefont {H.}~\bibnamefont {Schultheiss}},\
  }\bibfield  {title} {\enquote {\bibinfo {title} {Excitation of whispering
  gallery magnons in a magnetic vortex},}\ }\href@noop {} {\bibinfo  {journal}
  {ArXiv e-prints}\ }\BibitemShut {NoStop}%
\bibitem [{\citenamefont {K{\"o}rber}\ \emph {et~al.}(2020)\citenamefont
  {K{\"o}rber}, \citenamefont {Schultheiss}, \citenamefont {Hula},
  \citenamefont {Verba}, \citenamefont {Fa{\ss}bender}, \citenamefont
  {K{\'a}kay},\ and\ \citenamefont
  {Schultheiss}}]{korberNonlocalStimulationThreemagnon2020}%
  \BibitemOpen
\bibfield  {journal} {  }\bibfield  {author} {\bibinfo {author} {\bibfnamefont
  {L.}~\bibnamefont {K{\"o}rber}}, \bibinfo {author} {\bibfnamefont
  {K.}~\bibnamefont {Schultheiss}}, \bibinfo {author} {\bibfnamefont
  {T.}~\bibnamefont {Hula}}, \bibinfo {author} {\bibfnamefont {R.}~\bibnamefont
  {Verba}}, \bibinfo {author} {\bibfnamefont {J.}~\bibnamefont
  {Fa{\ss}bender}}, \bibinfo {author} {\bibfnamefont {A.}~\bibnamefont
  {K{\'a}kay}},\ and\ \bibinfo {author} {\bibfnamefont {H.}~\bibnamefont
  {Schultheiss}},\ }\bibfield  {title} {\enquote {\bibinfo {title} {Nonlocal
  stimulation of three-magnon splitting in a magnetic vortex},}\ }\href
  {https://doi.org/10.1103/PhysRevLett.125.207203} {\bibfield  {journal}
  {\bibinfo  {journal} {Physical Review Letters}\ }\textbf {\bibinfo {volume}
  {125}},\ \bibinfo {pages} {207203} (\bibinfo {year} {2020})}\BibitemShut
  {NoStop}%
\bibitem [{\citenamefont {Verba}\ \emph {et~al.}(2021)\citenamefont {Verba},
  \citenamefont {K{\"o}rber}, \citenamefont {Schultheiss}, \citenamefont
  {Schultheiss}, \citenamefont {Tiberkevich},\ and\ \citenamefont
  {Slavin}}]{verbaTheoryThreemagnonInteraction2021}%
  \BibitemOpen
  \bibfield  {author} {\bibinfo {author} {\bibfnamefont {R.}~\bibnamefont
  {Verba}}, \bibinfo {author} {\bibfnamefont {L.}~\bibnamefont {K{\"o}rber}},
  \bibinfo {author} {\bibfnamefont {K.}~\bibnamefont {Schultheiss}}, \bibinfo
  {author} {\bibfnamefont {H.}~\bibnamefont {Schultheiss}}, \bibinfo {author}
  {\bibfnamefont {V.}~\bibnamefont {Tiberkevich}},\ and\ \bibinfo {author}
  {\bibfnamefont {A.}~\bibnamefont {Slavin}},\ }\bibfield  {title} {\enquote
  {\bibinfo {title} {Theory of three-magnon interaction in a vortex-state
  magnetic nanodot},}\ }\href {https://doi.org/10.1103/PhysRevB.103.014413}
  {\bibfield  {journal} {\bibinfo  {journal} {Physical Review B}\ }\textbf
  {\bibinfo {volume} {103}},\ \bibinfo {pages} {014413} (\bibinfo {year}
  {2021})},\ \bibinfo {note} {comment: 10 pages, 4 figures},\ \Eprint
  {https://arxiv.org/abs/2008.11812} {arXiv:2008.11812} \BibitemShut {NoStop}%
\bibitem [{\citenamefont {Qin}, \citenamefont {H{\"a}m{\"a}l{\"a}inen},\ and\
  \citenamefont {{van
  Dijken}}(2018)}]{qinExchangetorqueinducedExcitationPerpendicular2018}%
  \BibitemOpen
  \bibfield  {author} {\bibinfo {author} {\bibfnamefont {H.}~\bibnamefont
  {Qin}}, \bibinfo {author} {\bibfnamefont {S.~J.}\ \bibnamefont
  {H{\"a}m{\"a}l{\"a}inen}},\ and\ \bibinfo {author} {\bibfnamefont
  {S.}~\bibnamefont {{van Dijken}}},\ }\bibfield  {title} {\enquote {\bibinfo
  {title} {Exchange-torque-induced excitation of perpendicular standing spin
  waves in nanometer-thick {{YIG}} films},}\ }\href
  {https://doi.org/10.1038/s41598-018-23933-y} {\bibfield  {journal} {\bibinfo
  {journal} {Scientific Reports}\ }\textbf {\bibinfo {volume} {8}},\ \bibinfo
  {pages} {5755} (\bibinfo {year} {2018})}\BibitemShut {NoStop}%
\bibitem [{\citenamefont {Tacchi}\ \emph {et~al.}(2019)\citenamefont {Tacchi},
  \citenamefont {Silvani}, \citenamefont {Carlotti}, \citenamefont {Marangolo},
  \citenamefont {Eddrief}, \citenamefont {Rettori},\ and\ \citenamefont
  {Pini}}]{tacchiStronglyHybridizedDipoleexchange2019}%
  \BibitemOpen
  \bibfield  {author} {\bibinfo {author} {\bibfnamefont {S.}~\bibnamefont
  {Tacchi}}, \bibinfo {author} {\bibfnamefont {R.}~\bibnamefont {Silvani}},
  \bibinfo {author} {\bibfnamefont {G.}~\bibnamefont {Carlotti}}, \bibinfo
  {author} {\bibfnamefont {M.}~\bibnamefont {Marangolo}}, \bibinfo {author}
  {\bibfnamefont {M.}~\bibnamefont {Eddrief}}, \bibinfo {author} {\bibfnamefont
  {A.}~\bibnamefont {Rettori}},\ and\ \bibinfo {author} {\bibfnamefont {M.~G.}\
  \bibnamefont {Pini}},\ }\bibfield  {title} {\enquote {\bibinfo {title}
  {Strongly hybridized dipole-exchange spin waves in thin {{Fe-N}}
  ferromagnetic films},}\ }\href {https://doi.org/10.1103/PhysRevB.100.104406}
  {\bibfield  {journal} {\bibinfo  {journal} {Physical Review B}\ }\textbf
  {\bibinfo {volume} {100}},\ \bibinfo {pages} {104406} (\bibinfo {year}
  {2019})}\BibitemShut {NoStop}%
\bibitem [{\citenamefont {K{\"o}rber}\ and\ \citenamefont
  {K{\'a}kay}(2021)}]{korberNumericalReverseEngineering2021}%
  \BibitemOpen
  \bibfield  {author} {\bibinfo {author} {\bibfnamefont {L.}~\bibnamefont
  {K{\"o}rber}}\ and\ \bibinfo {author} {\bibfnamefont {A.}~\bibnamefont
  {K{\'a}kay}},\ }\bibfield  {title} {\enquote {\bibinfo {title} {Numerical
  reverse engineering of general spin-wave dispersions: {{Bridge}} between
  numerics and analytics using a dynamic-matrix approach},}\ }\href
  {https://doi.org/10.1103/PhysRevB.104.174414} {\bibfield  {journal} {\bibinfo
   {journal} {Physical Review B}\ }\textbf {\bibinfo {volume} {104}},\ \bibinfo
  {pages} {174414} (\bibinfo {year} {2021})}\BibitemShut {NoStop}%
\bibitem [{\citenamefont {Rych{\l}y}\ \emph {et~al.}(2019)\citenamefont
  {Rych{\l}y}, \citenamefont {Tkachenko}, \citenamefont {K{\l}os},
  \citenamefont {Kuchko},\ and\ \citenamefont
  {Krawczyk}}]{rychlySpinWaveModes2019}%
  \BibitemOpen
  \bibfield  {author} {\bibinfo {author} {\bibfnamefont {J.}~\bibnamefont
  {Rych{\l}y}}, \bibinfo {author} {\bibfnamefont {V.~S.}\ \bibnamefont
  {Tkachenko}}, \bibinfo {author} {\bibfnamefont {J.~W.}\ \bibnamefont
  {K{\l}os}}, \bibinfo {author} {\bibfnamefont {A.}~\bibnamefont {Kuchko}},\
  and\ \bibinfo {author} {\bibfnamefont {M.}~\bibnamefont {Krawczyk}},\
  }\bibfield  {title} {\enquote {\bibinfo {title} {Spin wave modes in a
  cylindrical nanowire in crossover dipolar-exchange regime},}\ }\href
  {https://doi.org/10.1088/1361-6463/aaf2fc} {\bibfield  {journal} {\bibinfo
  {journal} {Journal of Physics D: Applied Physics}\ }\textbf {\bibinfo
  {volume} {52}},\ \bibinfo {pages} {075003} (\bibinfo {year}
  {2019})}\BibitemShut {NoStop}%
\bibitem [{\citenamefont {K{\"o}rber}, \citenamefont {K{\'e}zsm{\'a}rki},\ and\
  \citenamefont {K{\'a}kay}(2022)}]{korberModeSplittingSpin2022}%
  \BibitemOpen
  \bibfield  {author} {\bibinfo {author} {\bibfnamefont {L.}~\bibnamefont
  {K{\"o}rber}}, \bibinfo {author} {\bibfnamefont {I.}~\bibnamefont
  {K{\'e}zsm{\'a}rki}},\ and\ \bibinfo {author} {\bibfnamefont
  {A.}~\bibnamefont {K{\'a}kay}},\ }\bibfield  {title} {\enquote {\bibinfo
  {title} {Mode splitting of spin waves in magnetic nanotubes with discrete
  symmetries},}\ }\href@noop {} {\bibfield  {journal} {\bibinfo  {journal}
  {arXiv:2202.06601 [cond-mat]}\ } (\bibinfo {year} {2022})},\ \Eprint
  {https://arxiv.org/abs/2202.06601} {arXiv:2202.06601 [cond-mat]} \BibitemShut
  {NoStop}%
\bibitem [{\citenamefont {Kalinikos}\ and\ \citenamefont
  {Slavin}(1986)}]{kalinikosTheoryDipoleexchangeSpin1986}%
  \BibitemOpen
  \bibfield  {author} {\bibinfo {author} {\bibfnamefont {B.~A.}\ \bibnamefont
  {Kalinikos}}\ and\ \bibinfo {author} {\bibfnamefont {A.~N.}\ \bibnamefont
  {Slavin}},\ }\bibfield  {title} {\enquote {\bibinfo {title} {Theory of
  dipole-exchange spin wave spectrum for ferromagnetic films with mixed
  exchange boundary conditions},}\ }\href
  {https://doi.org/10.1088/0022-3719/19/35/014} {\bibfield  {journal} {\bibinfo
   {journal} {Journal of Physics C: Solid State Physics}\ }\textbf {\bibinfo
  {volume} {19}},\ \bibinfo {pages} {7013--7033} (\bibinfo {year}
  {1986})}\BibitemShut {NoStop}%
\bibitem [{Note2()}]{Note2}%
  \BibitemOpen
  \bibinfo {note} {Note, that the mean curvature $ \protect \mathcal
  {H}(\protect \bm {r})$ is also defined along the thickness because of that,
  as we discuss here only shells which consist of extrusions of curved surfaces
  along their normal direction. Thus the local curvature inside the volume of
  the shell is given by the surface curvature of the respective extruded
  surface.}\BibitemShut {Stop}%
\bibitem [{\citenamefont {Gallardo}\ \emph
  {et~al.}(2019{\natexlab{b}})\citenamefont {Gallardo}, \citenamefont
  {{Alvarado-Seguel}}, \citenamefont {Schneider}, \citenamefont
  {{Gonzalez-Fuentes}}, \citenamefont {{Rold{\'a}n-Molina}}, \citenamefont
  {Lenz}, \citenamefont {Lindner},\ and\ \citenamefont
  {Landeros}}]{gallardoSpinwaveNonreciprocityMagnetizationgraded2019}%
  \BibitemOpen
  \bibfield  {author} {\bibinfo {author} {\bibfnamefont {R.~A.}\ \bibnamefont
  {Gallardo}}, \bibinfo {author} {\bibfnamefont {P.}~\bibnamefont
  {{Alvarado-Seguel}}}, \bibinfo {author} {\bibfnamefont {T.}~\bibnamefont
  {Schneider}}, \bibinfo {author} {\bibfnamefont {C.}~\bibnamefont
  {{Gonzalez-Fuentes}}}, \bibinfo {author} {\bibfnamefont {A.}~\bibnamefont
  {{Rold{\'a}n-Molina}}}, \bibinfo {author} {\bibfnamefont {K.}~\bibnamefont
  {Lenz}}, \bibinfo {author} {\bibfnamefont {J.}~\bibnamefont {Lindner}},\ and\
  \bibinfo {author} {\bibfnamefont {P.}~\bibnamefont {Landeros}},\ }\bibfield
  {title} {\enquote {\bibinfo {title} {Spin-wave non-reciprocity in
  magnetization-graded ferromagnetic films},}\ }\href
  {https://doi.org/10.1088/1367-2630/ab0449} {\bibfield  {journal} {\bibinfo
  {journal} {New Journal of Physics}\ }\textbf {\bibinfo {volume} {21}},\
  \bibinfo {pages} {033026} (\bibinfo {year} {2019}{\natexlab{b}})}\BibitemShut
  {NoStop}%
\bibitem [{\citenamefont
  {Kostylev}(2014)}]{kostylevInterfaceBoundaryConditions2014}%
  \BibitemOpen
  \bibfield  {author} {\bibinfo {author} {\bibfnamefont {M.}~\bibnamefont
  {Kostylev}},\ }\bibfield  {title} {\enquote {\bibinfo {title} {Interface
  boundary conditions for dynamic magnetization and spin wave dynamics in a
  ferromagnetic layer with the interface {{Dzyaloshinskii-Moriya}}
  interaction},}\ }\href {https://doi.org/10.1063/1.4883181} {\bibfield
  {journal} {\bibinfo  {journal} {Journal of Applied Physics}\ }\textbf
  {\bibinfo {volume} {115}},\ \bibinfo {pages} {233902} (\bibinfo {year}
  {2014})}\BibitemShut {NoStop}%
\bibitem [{\citenamefont {Etesamirad}\ \emph {et~al.}(2021)\citenamefont
  {Etesamirad}, \citenamefont {Rodriguez}, \citenamefont {Bocanegra},
  \citenamefont {Verba}, \citenamefont {Katine}, \citenamefont {Krivorotov},
  \citenamefont {Tyberkevych}, \citenamefont {Ivanov},\ and\ \citenamefont
  {Barsukov}}]{Etesamirad2021}%
  \BibitemOpen
  \bibfield  {author} {\bibinfo {author} {\bibfnamefont {A.}~\bibnamefont
  {Etesamirad}}, \bibinfo {author} {\bibfnamefont {R.}~\bibnamefont
  {Rodriguez}}, \bibinfo {author} {\bibfnamefont {J.}~\bibnamefont
  {Bocanegra}}, \bibinfo {author} {\bibfnamefont {R.}~\bibnamefont {Verba}},
  \bibinfo {author} {\bibfnamefont {J.}~\bibnamefont {Katine}}, \bibinfo
  {author} {\bibfnamefont {I.~N.}\ \bibnamefont {Krivorotov}}, \bibinfo
  {author} {\bibfnamefont {V.}~\bibnamefont {Tyberkevych}}, \bibinfo {author}
  {\bibfnamefont {B.}~\bibnamefont {Ivanov}},\ and\ \bibinfo {author}
  {\bibfnamefont {I.}~\bibnamefont {Barsukov}},\ }\bibfield  {title} {\enquote
  {\bibinfo {title} {Controlling magnon interaction by a nanoscale switch},}\
  }\href {https://doi.org/10.1021/acsami.1c01562} {\bibfield  {journal}
  {\bibinfo  {journal} {ACS Applied Materials \& Interfaces}\ }\textbf
  {\bibinfo {volume} {13}},\ \bibinfo {pages} {20288} (\bibinfo {year}
  {2021})}\BibitemShut {NoStop}%
\bibitem [{\citenamefont {Tyberkevych}\ \emph {et~al.}()\citenamefont
  {Tyberkevych}, \citenamefont {Slavin}, \citenamefont {Artemchuk},\ and\
  \citenamefont {Rowlands}}]{Tyberkevych_ArXiv}%
  \BibitemOpen
  \bibfield  {author} {\bibinfo {author} {\bibfnamefont {V.}~\bibnamefont
  {Tyberkevych}}, \bibinfo {author} {\bibfnamefont {A.}~\bibnamefont {Slavin}},
  \bibinfo {author} {\bibfnamefont {P.}~\bibnamefont {Artemchuk}},\ and\
  \bibinfo {author} {\bibfnamefont {G.}~\bibnamefont {Rowlands}},\ }\href@noop
  {} {\enquote {\bibinfo {title} {{Vector Hamiltonian Formalism for Nonlinear
  Magnetization Dynamics}},}\ }\bibinfo {note} {ArXiv:2011.13562
  [cond-mat.mtrl-sci]}\BibitemShut {NoStop}%
\bibitem [{\citenamefont {Yuan}\ \emph {et~al.}(2022)\citenamefont {Yuan},
  \citenamefont {Sui}, \citenamefont {Kang},\ and\ \citenamefont
  {Jia}}]{Yuan_APL2022_twisted_magnon}%
  \BibitemOpen
  \bibfield  {author} {\bibinfo {author} {\bibfnamefont {S.}~\bibnamefont
  {Yuan}}, \bibinfo {author} {\bibfnamefont {C.}~\bibnamefont {Sui}}, \bibinfo
  {author} {\bibfnamefont {J.}~\bibnamefont {Kang}},\ and\ \bibinfo {author}
  {\bibfnamefont {C.}~\bibnamefont {Jia}},\ }\bibfield  {title} {\enquote
  {\bibinfo {title} {Electric readout of bloch sphere spanned by twisted magnon
  modes},}\ }\href {https://doi.org/10.1063/5.0085775} {\bibfield  {journal}
  {\bibinfo  {journal} {Appl. Phys. Lett.}\ }\textbf {\bibinfo {volume}
  {120}},\ \bibinfo {pages} {132402} (\bibinfo {year} {2022})}\BibitemShut
  {NoStop}%
\bibitem [{\citenamefont {Gallardo}, \citenamefont {{Alvarado-Seguel}},\ and\
  \citenamefont {Landeros}(2022)}]{gallardoHighSpinwaveAsymmetry2022}%
  \BibitemOpen
  \bibfield  {author} {\bibinfo {author} {\bibfnamefont {R.~A.}\ \bibnamefont
  {Gallardo}}, \bibinfo {author} {\bibfnamefont {P.}~\bibnamefont
  {{Alvarado-Seguel}}},\ and\ \bibinfo {author} {\bibfnamefont
  {P.}~\bibnamefont {Landeros}},\ }\bibfield  {title} {\enquote {\bibinfo
  {title} {High spin-wave asymmetry and emergence of radial standing modes in
  thick ferromagnetic nanotubes},}\ }\href
  {https://doi.org/10.1103/PhysRevB.105.104435} {\bibfield  {journal} {\bibinfo
   {journal} {Physical Review B}\ }\textbf {\bibinfo {volume} {105}},\ \bibinfo
  {pages} {104435} (\bibinfo {year} {2022})}\BibitemShut {NoStop}%
\bibitem [{\citenamefont {Guslienko}\ and\ \citenamefont
  {Slavin}(2000)}]{Guslienko_JAP2000}%
  \BibitemOpen
  \bibfield  {author} {\bibinfo {author} {\bibfnamefont {K.~Y.}\ \bibnamefont
  {Guslienko}}\ and\ \bibinfo {author} {\bibfnamefont {A.~N.}\ \bibnamefont
  {Slavin}},\ }\bibfield  {title} {\enquote {\bibinfo {title} {{Spin-waves in
  cylindrical magnetic dot arrays with in-plane magnetization}},}\ }\href
  {https://doi.org/10.1063/1.372698} {\bibfield  {journal} {\bibinfo  {journal}
  {J. Appl. Phys.}\ }\textbf {\bibinfo {volume} {87}},\ \bibinfo {pages} {6337}
  (\bibinfo {year} {2000})}\BibitemShut {NoStop}%
\bibitem [{\citenamefont {Guslienko}\ and\ \citenamefont
  {Slavin}(2011)}]{Guslienko_JMMM2011}%
  \BibitemOpen
  \bibfield  {author} {\bibinfo {author} {\bibfnamefont {K.~Y.}\ \bibnamefont
  {Guslienko}}\ and\ \bibinfo {author} {\bibfnamefont {A.~N.}\ \bibnamefont
  {Slavin}},\ }\bibfield  {title} {\enquote {\bibinfo {title} {Magnetostatic
  green's functions for the description of spin waves in finite rectangular
  magnetic dots and stripes},}\ }\href
  {https://doi.org/10.1016/j.jmmm.2011.05.020} {\bibfield  {journal} {\bibinfo
  {journal} {J. Magn. Magn. Mater.}\ }\textbf {\bibinfo {volume} {323}},\
  \bibinfo {pages} {2418} (\bibinfo {year} {2011})}\BibitemShut {NoStop}%
\bibitem [{\citenamefont {Arfken}\ and\ \citenamefont
  {Weber}(2001)}]{Arfken_Book}%
  \BibitemOpen
  \bibfield  {author} {\bibinfo {author} {\bibfnamefont {G.~B.}\ \bibnamefont
  {Arfken}}\ and\ \bibinfo {author} {\bibfnamefont {H.~J.}\ \bibnamefont
  {Weber}},\ }\href@noop {} {\emph {\bibinfo {title} {{Mathematical Methods for
  Physicists}}}},\ \bibinfo {edition} {5th}\ ed.\ (\bibinfo  {publisher}
  {Academic Press},\ \bibinfo {year} {2001})\BibitemShut {NoStop}%
\end{thebibliography}
%aipnum4-2.bst 2019-01-14 (MD) hand-edited version of apsrev4-1.bst

\end{document}